\documentclass[11pt,psfig]{article}
\usepackage{apacite}
 
\usepackage{graphicx}
 \usepackage{amsmath}
 \usepackage{amssymb}
\usepackage{color}
\usepackage{ulem}
\usepackage{placeins}
\usepackage{leftidx}

\usepackage{xcolor,colortbl}
\definecolor{Gray}{gray}{0.85}
\newcolumntype{g}{>{\columncolor{Gray}}c}
\newcolumntype{w}{>{\columncolor{white}}c}
\definecolor{LightCyan}{rgb}{0.88,1,1}

\newcommand{\tb}{\textcolor{black}}

\newcommand{\tc}{\textcolor{black}}
\newcommand{\tnr}{\textcolor{black}}
\newcommand{\tr}{\textcolor{black}}

\newcommand{\tor}{\textcolor{black}}

\newcommand{\tnb}{\textcolor{black}}
\newcommand{\tbl}{\textcolor{black}}

\newcommand{\Var}{{\rm Var\,}}

\newcommand{\var}{\Var}

\newtheorem{Remark}{Remark}[section]

\newtheorem{Algorithm}{Algorithm}[section]

  \title{Predictive inference for locally stationary time series \tc{with an application to climate data}
}

\author{Srinjoy Das \\
Department of Electrical \\ and Computer Engineering\\
     University of California---San Diego \\
   La Jolla, CA 92093, USA 
   \\email: {\tt   s2das@ucsd.edu}
\and
Dimitris N. Politis 
\\Department of Mathematics \\
     University of California---San Diego \\
   La Jolla, CA 92093-0112, USA 
  \\   email: {\tt   dpolitis@ucsd.edu} 
 }
\date{ } 

\begin{document}

\newcolumntype{g}{>{\columncolor{Gray}}c}
 \maketitle

\begin{abstract}
The Model-free Prediction Principle of Politis \tnr{(2015)} has been 
successfully applied to \tnr{general} regression problems, as well as 
problems involving stationary time series. However, 
with long time series, e.g. annual temperature measurements spanning over 100 years
or daily financial returns spanning several years, it  may be unrealistic to assume stationarity 
throughout the span of the dataset. In the paper at hand, we show how Model-free Prediction
can be applied to handle time series that are only locally  stationary,
i.e., they can be \tnr{assumed to be} as stationary only over short 
time-windows. \tnr{Surprisingly there is little literature on point prediction for general locally stationary time series even in model-based setups and there is no literature on the construction of prediction intervals of locally stationary time series. We attempt to fill this gap here as well.} Both one-step-ahead point  predictors and prediction intervals 
are constructed, and \tb {the performance of model-free is compared to model-based prediction using}
models that incorporate a trend
and/or heteroscedasticity. \tb {Both aspects of the paper, model-free and model-based, are novel
in the context of time-series that are locally (but not globally) stationary.} \tc{We also demonstrate the application of our Model-based
and Model-free prediction methods to speleothem climate data which exhibits local stationarity \tnr{and show} that our best \tnr{model-free} point prediction results outperform that obtained with the RAMPFIT algorithm previously used for analysis of this data.}
\end{abstract}


\newpage

  \section{Introduction}

Consider a real-valued time series   
dataset    $Y_1, \ldots, Y_n $
spanning a   long time interval, e.g. annual temperature measurements spanning over 100 years
or daily financial returns spanning several years.
 It may be unrealistic to assume that 
the stochastic structure of time series $\{Y_t , t\in {\bf Z}  \}$ has stayed  invariant over such a  long stretch of time; hence, we can not
assume that $\{Y_t    \}$ is stationary. 
More realistic is to   assume a slowly-changing 
stochastic structure, i.e.,  a {\it locally  stationary model} -- see
\tr {\cite{priestley1965evolutionary}, \cite{priestley1988non}, \cite{dahlhaus1997fitting} and \cite{dahlhaus2012locally}.}
 
Our objective is predictive inference for the next 
data point $Y_{n+1}$, i.e., constructing a point and interval predictor for $Y_{n+1}$.
The usual approach  for dealing with nonstationary series is to assume 
that the data can be decomposed as the sum of three components:
$$  
 \mu( t)+ S_t + W_t 
$$  
where $\mu( t)$ is a deterministic trend function, $S_t$ is a seasonal (periodic) time series, 
and $\{W_t\}$  is (strictly)  stationary with mean zero;
this is   
the `classical' decomposition of a time series to trend, seasonal and stationary components.   
The seasonal (periodic) component, be it random or deterministic, can be easily estimated and
removed; see e.g.
\tr {\cite{brockwell2013time}}.
Having done that, the `classical' decomposition
simplifies to the following model with additive trend, i.e.,
\begin{equation}
Y_t=\mu( t)+   W_t
 \label{NSTS.eq.model homo}
\end{equation}
which 
 can be generalized to accomodate
a     time-changing variance as well, i.e., 
 \begin{equation}
Y_t=\mu( t)+ \sigma (t) W_t . 
 \label{NSTS.eq.model hetero}
\end{equation}
In both above models, the time series $\{W_t\}$
  is assumed to be (strictly)  stationary,
weakly dependent, e.g. strong mixing, and satisfying $EW_t =0$;
in model \eqref{NSTS.eq.model hetero}, it is also
assumed that    $\var(W_t)=1$.  
As usual, the deterministic functions 
 $\mu( \cdot)$ and  $\sigma (\cdot)$ are unknown but
assumed to belong to 
 a class of functions  that is either finite-dimensional (parametric) or not \tr {(nonparametric)};
we will focus on the latter, in which case
it is customary to assume that $\mu( \cdot)$ and  $\sigma (\cdot)$  possess some degree of smoothness, i.e., that  $\mu(  t)$ and  $\sigma ( t)$ change
smoothly (and slowly) with $t$.

\begin{Remark} [Quantifying smoothness] \rm
To analyze locally  stationary   series it is sometimes useful to
map the index set $\{1, \ldots, n\}$ onto the interval $[0,1]$. 
In that respect, consider two functions $  \mu_{_{[0,1]}} : [0,1] \mapsto {\bf R} $ and 
 $  \sigma_{_{[0,1]}}  : [0,1] \mapsto (0,\infty)$, and let 
\begin{equation}
\mu( t)=   \mu_{_{[0,1]}} (a_t) \ \mbox{ and } \ \sigma (t)=  \sigma_{_{[0,1]}} (a_t)
\label{NSTS.eq.qs}
\end{equation}
where 
$a_t=(t-1)/n$ for $t=1, \ldots, n$.
We will assume that $  \mu_{_{[0,1]}}( \cdot)$ and  $ \sigma_{_{[0,1]}} (\cdot)$
 are continuous and smooth, i.e.,   possess $k$ continuous derivatives 
on $[0,1]$. To take full advantage of the local linear
smoothers of Section \ref{NSTS.sec.trend} ideally one would need $k\geq 2$.
However, all methods to be discussed here are valid even when  $ \mu_{_{[0,1]}}( x)$ and  $  \sigma_{_{[0,1]}} (x)$  are continuous for all $x\in [0,1]$  but only piecewise smooth.  
\label{NSTS.re.smooth} 
\end{Remark}

As far as capturing the  first two moments of $Y_t$, 
models \eqref{NSTS.eq.model homo} and  \eqref{NSTS.eq.model hetero} 
are considered general and
 flexible---especially when $\mu( \cdot)$ and  $\sigma (\cdot)$ 
are not parametrically specified---and have been studied 
extensively; see e.g. 
\tr {\cite{zhou2009local}, \cite{zhou2010simultaneous}.}
However,   it may be  that the skewness and/or kurtosis of $Y_t$   changes with $ t$,
in which case   centering and studentization alone can not render the problem stationary. To see why, note that  
under model  \eqref{NSTS.eq.model hetero},
$ EY_t=\mu( t)$ and $ \Var Y_t=\sigma ^2( t)$; hence, 
\begin{equation}
W_t =\frac{Y_t-\mu( t)}{ \sigma (t)}      
\label{NSTS.eq.W}
\end{equation}
  cannot be (strictly) stationary unless 
the skewness and  kurtosis of $Y_t$ are constant. 
Furthermore, it may be the case that the nonstationarity is due to 
a feature of the $m$--th dimensional marginal distribution not being constant
for some $m\geq 1$, 
e.g., perhaps the correlation Corr$(Y_t, Y_{t+1})$ 
changes smoothly (and slowly) with $t$. Notably, models \eqref{NSTS.eq.model homo} and  \eqref{NSTS.eq.model hetero} only concern themselves with features of the 
1st marginal distribution.

For all the above reasons, it seems valuable to develop a methodology for
the statistical analysis of nonstationary time series that does not
rely on  simple additive models such as   \eqref{NSTS.eq.model homo} and  \eqref{NSTS.eq.model hetero}. Fortunately, the 
Model-free Prediction Principle of \tb {\cite{Politis2013}, \cite{politis2015model}}
 \tb{suggests a way} to accomplish  Model-free inference---including
the  construction of  prediction intervals---in the general setting of
   time series that are only locally stationary.
The key towards Model-free inference is to be able to construct an invertible transformation 
$H_n: \underline{Y}_n \mapsto \underline \epsilon_{n}$
where $\underline \epsilon_{n}=(\epsilon_{1}, \ldots, \epsilon_{n} )'$
is  a random vector with i.i.d.~components;
the details are given in Section \ref{NSTS.Model-free inference}.
The next section revisits the problem of model-based  
inference in a locally stationary setting, and develops
a bootstrap methodology for the construction of 
(model-based) prediction intervals. 
Both approaches, Model-based \tb {of Section \ref{NSTS.Model-based inference}} and Model-free \tb{of Section \ref{NSTS.Model-free inference}}, are novel, and they are empirically compared to each other in Section \ref{NSTS.Numerical} \tb {using finite sample \tor{experiments}}. Both synthetic and real-life data are used for this purpose. 

\tc{The prototype of local (but not global) stationarity is \tnr{manifested in} climate data observed over long periods. In Section \ref{NSTS.Speleothem} we focus on the speleothem climate archive data discussed in \cite{fleitmann2003holocene} whose statistical analysis is presented in \cite{mudelsee2014climate}.
This dataset which is shown in Figure \ref{Mudelsee_full} contains oxygen isotope record obtained from stalagmite Q5 from southern Oman over the past 10,300 years. In this figure \tb{delta-O-18} on the Y-axis is a measure of the ratio of stable isotopes oxygen-18 ($^{18}O$) and  oxygen-16 ($^{16}O$) and \tb{Age (a B.P. where B.P. indicates Before Present)} on the X-axis denotes time before the present i.e. time increases from right to left. Details of how \tb{delta-O-18} is defined can be found on \url{https://en.wikipedia.org/wiki/\%CE\%9418O}. 
Along the growth axis of the nearly 1 meter long speleothem (which is in this case stalagmite), approximately every 0.7 mm about 5 mg material (calcium carbonate) was drilled, thereby yielding n=1345 samples. This carbonate was then analyzed to determine the \tb{delta-O-18} values.}

The oxygen isotope ratio serves as a proxy variable for the climate variable {\bf monsoon rainfall}. This data can be used for climate analysis applications such as whether there exists solar influences on the variations in monsoon rainfall; here low values of delta-O-18 would indicate a strong monsoon. The full dataset can be referenced at:\\
\url{http://manfredmudelsee.com/book/data/1-7.txt}.
 Previously the RAMPFIT algorithm \cite{mudelsee2000ramp} has been used to fit data that exhibit change points such as the speleothem climate archive. \tnb{However RAMPFIT was not designed to handle arbitrary locally stationary data which maybe present in  climate time series. In Section \ref{NSTS.Speleothem} we focus on a part of the delta-O-18 proxy variable data that contains a linear trend and apply our Model-Free and Model-Based algorithms over this range to estimate the performance of both point prediction and prediction intervals. We then show that our best Model-Free point predictor achieves superior performance in point prediction compared to RAMPFIT; notably, RAMPFIT was not originally designed to estimate prediction intervals.}

In Section \ref{NSTS.Diagnostics} we also describe \tnr{techniques for diagnostics} which \tor{are useful for} Model-Free prediction in order to successfully generate both point predictors and prediction intervals. Model-Based and Model-Free algorithms for the construction of prediction intervals are described in detail in Appendix \ref{Appendix_Boot}. \tnr{The RAMPFIT algorithm used to generate point prediction results for comparison with our model-free and model-based methods is described in Appendix \ref{Appendix_Rampfit}}.

\graphicspath{{/Users/rumpagiri/Documents/NONPARAMETRIC/model_free/papers/LSTS}}
\DeclareGraphicsExtensions{.png}

{\begin{figure}[!t]
  \centering
  \includegraphics[width=3.5in, height=2.5in]{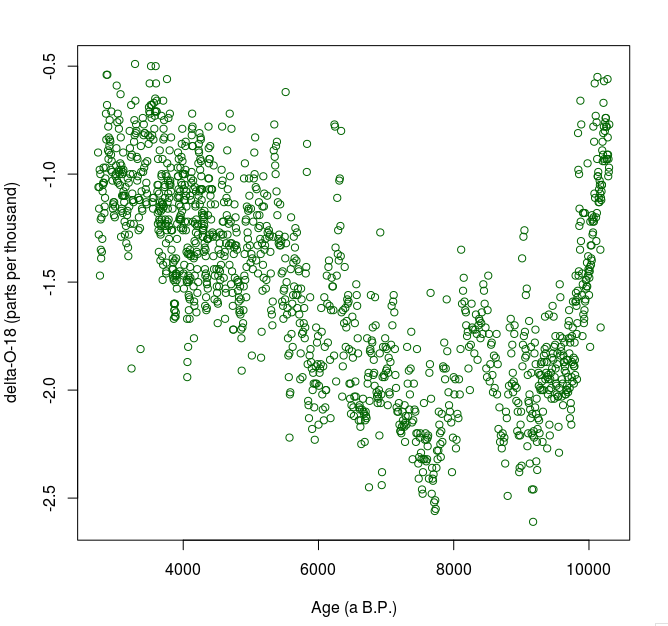}
  \caption{Oxygen Isotope Record from stalagmite Q5 from southern Oman (1345 samples) where B.P. indicates Before Present}
  \label{Mudelsee_full}
\end{figure}}

 

\section{Model-based inference}
\label{NSTS.Model-based inference}

Throughout Section \ref{NSTS.Model-based inference}, we will assume  
model   \eqref{NSTS.eq.model hetero}---that
includes model  \eqref{NSTS.eq.model homo} as a special case---together with
a nonparametric assumption on smoothness of $\mu(\cdot)$ and  $\sigma ( \cdot)$ as
described in Remark \ref{NSTS.re.smooth}.
   
\subsection{Theoretical optimal point prediction}
\label{NSTS.sec.TOPP}

It is well-known that the $L_2$--optimal 
predictor of  $Y_{n+1}$ given the data $\underline{Y}_n=(Y_1,\ldots, Y_n)'$ is 
the conditional expectation $E(Y_{n+1}|\underline{Y}_n)$.
Furthermore, under model  \eqref{NSTS.eq.model hetero}, we have 
\begin{equation}
E(Y_{n+1}|\underline{Y}_n)=\mu( n+1)+ \sigma (n+1) E(W_{n+1}|\underline{Y}_n).
\label{NSTS.eq.point pred}
\end{equation}

For \tr {$j<J$}, define ${\cal F}_j^J(Y)$ to be the
{\it information set}   $\{ Y_j, Y_{j+1},\ldots, Y_J\}$,
also known as     $\sigma$--field, 
and note that the information sets 
${\cal F}_{-\infty}^t(Y)$ and ${\cal F}_{-\infty}^t(W)$ are identical for any $t$,
i.e., knowledge of  $\{ Y_s$ for $s<t\}$ is equivalent to 
knowledge of { 
\tr {$\{ W_s$ for $s<t\}$}}; here, $\mu(\cdot)$ and  $\sigma ( \cdot)$
are assumed known. 
Hence, for large $n$, and due to the assumption that $W_t$
is weakly dependent (and therefore
  the same must be true for $Y_t$ as well), the following large-sample 
approximation is useful, i.e., 
\begin{equation} 
E(W_{n+1}|\underline{Y}_n)  \simeq  E(W_{n+1}|Y_s, s\leq n) =
E(W_{n+1}|W_s, s\leq n)  \simeq E(W_{n+1}|\underline{W}_n )
\label{NSTS.eq.point pred approx}
\end{equation}
where $\underline{W}_n = (W_1, \ldots, W_n)'$.

All that is needed now is to construct an approximation for 
$E(W_{n+1}|\underline{W}_n )$. Usual approaches involve either
 assuming that the time series $\{W_t\}$ is Markov of order $p$ as in 
\tr {\cite{pan2016bootstrap}},
or   approximating $E(W_{n+1}|\underline{W}_n )$ by a linear 
function of $\underline{W}_n$ as in 
\tr {\cite{mcmurry2015high}},
i.e., contend ourselves with the best linear predictor of $W_{n+1}$
denoted by  $\bar E(W_{n+1}|\underline{W}_n )$.

Taking the latter approach,  
the $L_2$--optimal linear predictor of $W_{n+1}$  based on $\underline{W}_n $ is 
\begin{equation} \label{NSTS.eq.opt}
 \bar E(W_{n+1}|\underline{W}_n ) = \phi_{1}(n) W_n + \phi_{2}(n) W_{n-1} + \ldots + \phi_{n}(n) W_1,
 \end{equation}
where the optimal coefficients $\phi_{i}(n)$ are computed from the normal 
equations, i.e.,  
$ \phi(n) \equiv (\phi_{1}(n) , \cdots, \phi_{n}(n))'= \Gamma_n^{-1}  \gamma(n)$; 
here, $\Gamma_n = [\gamma_{|i-j|}]_{i,j=1}^n$ is the autocovariance matrix of
the  random vector $\underline{W}_n $, and $ \gamma(n) = (\gamma_1, \ldots, \gamma_n)'$  where $\gamma_k=EY_j Y_{j+k} $. 
Of course,  $\Gamma_n$ is unknown but can be estimated by any of the positive
definite estimators developed in 
\tr {\cite{mcmurry2015high}}.

Alternatively, the $L_2$--optimal linear predictor of $W_{n+1}$ can be obtained by fitting
a (causal) AR($p$) model to the data $W_1, \ldots, W_n$ with $p$ chosen by minimizing AIC or a related criterion; this would entail fitting the model:
\begin{equation} \label{NSTS.eq.AR}
 W_t  = \phi_{1}  W_{t-1} + \phi_{2}  W_{t-2} + \cdots + \phi_{p}  W_{t-p}+ V_t
 \end{equation}
where $V_t$ is a  stationary white noise,
i.e., an uncorrelated sequence, with mean zero and variance~$\tau^2$.
The   implication then is that 
\begin{equation} \label{NSTS.eq.ARpredictor}
\bar E(W_{n+1}|\underline{W}_n ) = \phi_{1}  W_n + \phi_{2}  W_{n-1} + \cdots + \phi_{p}  W_{n-p+1}.
\end{equation} 
As discussed in 
\tr {the rejoinder to \cite{mcmurry2015high}},
the two methods for constructing $\bar E(W_{n+1}|\underline{W}_n ) $ are closely related; in fact,   predictor \eqref{NSTS.eq.opt}   coincides with the above AR--type predictor if the matrix    $\Gamma_n$   is the one implied by the
fitted AR($p$) model \eqref{NSTS.eq.AR}.
We will use the AR--type predictor in the sequel because it additionally
affords us the possibility of resampling based on model~\eqref{NSTS.eq.AR}.

\subsection{Trend estimation and practical prediction}
\label{NSTS.sec.trend}

To construct the $L_2$--optimal predictor \eqref{NSTS.eq.point pred}, 
we need to estimate  the smooth trend $\mu(\cdot)$ and
variance   $\sigma ( \cdot)$
in a nonparametric fashion; this can be easily accomplished via kernel smoothing---see 
  e.g. 
\tr {\cite{hardle1992kernel}, \cite{kim1996bandwidth}, \cite{li2007nonparametric}}.
When confidence intervals for $\mu( t)$ and  $\sigma (  t)$ 
are required, however,  matters are more complicated as the asymptotic distribution
of the different estimators depends on many unknown parameters; see e.g. 
 \tr {\cite{masry1995nonparametric}}.
Even more difficult is the construction of prediction intervals.

Note, furthermore, that the problem of prediction of $Y_{n+1}$ involves
estimating the functions $  \mu_{_{[0,1]}}(a)$ and  $  \sigma_{_{[0,1]}} ( a)$  
described in Remark \ref{NSTS.re.smooth} for $a=1$, i.e., it 
is essentially a boundary problem. In such cases, it is well-known that
  local linear   fitting has better properties---in particular, smaller bias---than
kernel smoothing which is well-known to be tantamount to local constant fitting;
\tr{\cite{fan1996local},\cite{fan2007nonlinear}, or \cite{li2007nonparametric}}.

\begin{Remark} [One-sided estimation] \rm
Since the goal is predictive inference on $Y_{n+1}$,
local constant   and/or  local linear fitting must be performed in 
a {\it one-sided way}. To see why, recall that in predictor \eqref{NSTS.eq.point pred}, the estimands involve $  \mu_{_{[0,1]}}(1)$ and  $ \sigma_{_{[0,1]}} ( 1)$
as just mentioned. \tb{Furthermore} to compute $\bar E(W_{n+1}|\underline{W}_n )$
in eq.~\eqref{NSTS.eq.opt} we need access to the stationary data $W_1,\ldots,W_n$
in order to estimate $\Gamma_n$. 
The $W_t$'s are not directly observed, but---much like residuals in a
regression---they can be reconstructed by eq.~\eqref{NSTS.eq.W} 
with estimates of $\mu(t)$ and   $\sigma (  t)$ plugged-in. 
What is important is that   {\bf the way $W_t$ is reconstructed/estimated  
by (say) $\hat W_t$ must remain the  same for all $t$}, otherwise 
the reconstructed data $\hat W_1,\ldots,\hat W_n$ can not be considered stationary.
Since $W_t$ can only be estimated in a one-sided way for $t$ close to $n$,
the same one-sided way must also be implemented for  $t$ in the middle of 
the dataset even though in that case two-sided estimation is possible.    
\label{NSTS.re.onesided}
\end{Remark}
 
\tb{By}  analogy to model-based regression 
\tr {as described in \cite{Politis2013}},
the one-sided Nadaraya-Watson (NW) kernel estimators of $\mu(t)$ and $\sigma ( t )$
can be defined in two ways. In what follows, the notation
$t_k=k$ will be used; this may appear redundant but
it makes clear that $t_k$ is the $k$th design point in the
time series regression, and  allows for easy 
extension   in the case of missing   data. 
Note that the bandwidth parameter $b$
  will be assumed to satisfy 
\begin{equation}
b\to \infty \ \mbox{as} \ n\to \infty \ \mbox{but} \ b/n\to 0,
\label{NSTS.eq.new bandwidth}
\end{equation}
 i.e., $b$ is analogous to the product $hn$ \tb{where $h$ is the usual bandwidth in nonparametric regression, see e.g.}
We will assume throughout that $K(\cdot)$  is a nonnegative, symmetric kernel function.
\begin{enumerate}

\item  {\bf NW--Regular fitting:}   
Let $t\in [b+1, n]$, and define 

\begin{equation}
\hat \mu(t) = \sum _{i=1}^{t } \ Y_{i} \ \hat K\left(\frac{t-t_{i}}{b}\right)
\ \ \mbox{and} \ \ 
\hat M(t) = \sum _{i=1}^{t } \ Y_{i}^2 \ \hat K(\frac{t-t_{i}}{b})
\label{NSTS.eq.nw-mu}
\end{equation}
where
\begin{equation}
 \hat \sigma(t) = \sqrt { \hat M_{t} - \hat \mu(t)^2 }
\ \ \mbox{and} \ \ 
\hat K \left( \frac { t-t_{i} } {b}\right) = \frac { K(\frac{t-t_{i}} {b}) }{\sum _{k=1}^{t } K(\frac {t-t_{k}}{b})} .  
\label{NSTS.eq.nw-sigma}
\end{equation}
 
Using $\hat \mu(t)$ and $\hat \sigma(t)$ we can now define the
{\it fitted} residuals by 
\begin{equation}
\hat W_t= \frac{Y_t- \hat \mu(t)}{ \hat \sigma (t )}
\ \ \mbox{for} \ \ 
t=b+1,\ldots,n.
\label{NSTS.eq.hatW}
\end{equation}
 
\item  {\bf NW--Predictive   fitting (delete-1):}  
Let
\begin{equation}
\tilde \mu(t) = \sum _{i=1}^{t-1} \ Y_{i} \ \tilde K\left(\frac{t-t_{i}}{b}\right)
\ \ \mbox{and} \ \ 
\tilde M(t) = \sum _{i=1}^{t-1} \ Y_{i}^2 \ \tilde K(\frac{t-t_{i}}{b})
\label{NSTS.eq.nw-muPRED}
\end{equation}
where
\begin{equation}
 \tilde \sigma(t) = \sqrt { \tilde M_{t} - \tilde \mu(t)^2 }
\ \ \mbox{and} \ \ 
\tilde K \left( \frac { t-t_{i} } {b}\right) = \frac { K(\frac{t-t_{i}} {b}) }{\sum _{k=1}^{t-1} K(\frac {t-t_{k}}{b})}   .
\label{NSTS.eq.nw-sigmaPRED}
\end{equation}
Using $\tilde \mu(t)$ and $\tilde \sigma(t)$ we  now define the
{\it predictive} residuals by 
\begin{equation}
\tilde W_t= \frac{Y_t- \tilde \mu(t)}{ \tilde \sigma (t )}
\ \ \mbox{for} \ \ 
t=b+1,\ldots,n.
\label{NSTS.eq.tildeW}
\end{equation}

\end{enumerate}
 \vskip .1in
 \noindent
Similarly, the one-sided local linear (LL) fitting   estimators of $\mu(t)$ and $\sigma ( t )$
can be defined in two ways.

\begin{enumerate}

\item  {\bf LL--Regular fitting:}   
Let $t\in [b+1, n]$, and define 
\begin{equation}
\hat \mu(t)=\frac{ \sum_{j=1}^{t } w_jY_j }{\sum_{j=1}^{t } w_j + n^{-2}}
\ \ \mbox{and} \ \ 
\hat M(t) = \frac{ \sum_{j=1}^{t } w_jY_j^2 }{\sum_{j=1}^{t } w_j + n^{-2}}
\label{NSTS.eq.locallinearF}
\end{equation}
where 
\begin{equation}
w_j= K(\frac{t-t_{j}} {b}) \left[ s_{t,2} -(t-t_{j})s_{t,1}
\right],
\label{NSTS.eq.locallinearweightsF}
\end{equation}
and
$ s_{t,k} =\sum_{j=1}^{t }K(\frac{t-t_{j}} {b}) (t-t_{j})^k$
  for $k=0,1,2$.
The term $ n^{-2}$  in eq.~\eqref{NSTS.eq.locallinearF} is just 
to ensure   the denominator is not zero; see    Fan  (1993).
Eq.~\eqref{NSTS.eq.nw-sigma} then  yields $\hat \sigma(t)$,
and eq.~\eqref{NSTS.eq.hatW} yields~$\hat W_t$.

\item  {\bf LL--Predictive   fitting (delete-1):}  
Let
 \begin{equation}
\tilde \mu(t)=\frac{ \sum_{j=1}^{t-1 } w_jY_j }{\sum_{j=1}^{t-1 } w_j + n^{-2}}
\ \ \mbox{and} \ \ 
\tilde  M(t) =\frac{ \sum_{j=1}^{t-1 } w_jY_j^2 }{\sum_{j=1}^{t-1 } w_j + n^{-2}}
\label{NSTS.eq.locallinearP}
\end{equation}
where 
\begin{equation}
w_j= K(\frac{t-t_{j}} {b}) \left[ s_{t-1,2} -(t-t_{j})s_{t-1,1}
\right].
\label{NSTS.eq.locallinearweightsF}
\end{equation}
Eq.~\eqref{NSTS.eq.nw-sigmaPRED} then  yields $\tilde \sigma(t)$,
and eq.~\eqref{NSTS.eq.tildeW} yields $\tilde W_t$.

\end{enumerate}

\vskip .1in

\noindent
Using one of the above four methods (NW vs.~LL, regular vs.~predictive)
gives estimates of the quantities needed to compute the $L_2$--optimal 
predictor \eqref{NSTS.eq.point pred}. In order to 
approximate $E(W_{n+1}|\underline{Y}_n)$, one would treat the proxies  
$\hat W_t$ or $\tilde W_t$ as if they were the true $W_t$, and proceed
as outlined in Section \ref{NSTS.sec.TOPP}.

\begin{Remark} [Predictive vs.~regular fitting]
\label{NSTS.re.Predictive vs. regular fitting}     \rm
In order 
to estimate $  \mu (n+1)$ and $  \sigma (n+1)$, the predictive 
fits $\tilde \mu (n+1)$ and $\tilde \sigma (n+1)$ are 
constructed in a straightforward manner. However, the 
formula giving $\hat \mu (t)$ and $\hat \sigma (t)$ 
changes when $t$ becomes greater than $n$;  this is due to an  
effective change in kernel shape since part of the kernel is not used
when $t>n$. Focusing momentarily on the trend estimators,
what happens is that the formulas for $\tilde \mu (t)$
and $\hat  \mu (t)$---although different when  
$t\leq n$---become  identical when $t>n$  except for the difference in 
kernel shape. 
Traditional model-fitting   ignores these issues, i.e., proceeds
with using   different formulas for estimation of $\mu (t)$
according to whether 
$t\leq n$ or $t>n$. 
However, in trying to predict the new, unobserved $W_{n+1}$ we need
to first capture its statistical characteristics, and for this reason we need
a sample of $W_t$'s. But the residual from the model 
at $t=n+1$ looks like $\tilde W_{n+1}$ from {\it either}
regular or predictive  approach, since $\tilde \mu (t)$
and $\hat  \mu (t)$ 
become the same when $t=n+1$;   it is apparent that traditional model-fitting  tries to capture the statistical characteristics of 
$\tilde W_{n+1}$ from a sample of $\hat W_t$'s, i.e., comparing apples to 
oranges. Herein lies the 
problem which is analogous to the discussion on prediction using
fitted vs.~predictive residuals in nonparametric 
regression
\tr {as discussed in \cite{Politis2013}}.
Therefore, our preference is to use
the predictive quantities  $\tilde \mu (t)$, $\tilde \sigma (t)$,
and $\tilde W_t$ throughout  the predictive modeling.
\end{Remark}

 \begin{Remark} [Time series  cross-validation] \rm
 To choose the  bandwidth $b$ for either of the above methods,
 predictive cross-validation   may be used
but  it must be adapted to the time series prediction setting, i.e., always 
one-step-ahead. To elaborate, let  $k<n$, and suppose only 
subseries $Y_1,\ldots, Y_k$ has been observed.
Denote
$\hat Y_{k+1}$ the best predictor of $Y_{k+1}$ based on the
data $Y_1,\ldots, Y_k$  constructed according to the
above methodology and some choice of $b$. However, since $Y_{k+1}$ is known, the
quality of the predictor can be assessed. So, for each value of $b$
over a reasonable range,  we can form either
 $PRESS(b)=\sum_{k=k_o}^{n-1} ( \hat Y_{k+1} - Y_{k+1})^2 $
or  $PRESAR(b)=\sum_{k=k_o}^{n-1} | \hat Y_{k+1} - Y_{k+1}| $; here
$k_o$ should be big enough so that estimation is accurate,
e.g., $k_o$ can be of the order of $\sqrt{n}$.
The cross-validated  bandwidth choice would then be the $b$ that
minimizes $PRESS(b)$; alternatively, we can choose to minimize $PRESAR(b)$ if
 an   $L_1$ measure of loss is preferred.
Finally, note that a quick-and-easy (albeit suboptimal) version of the above is to
use the (supoptimal) predictor  $\hat Y_{k+1}\simeq \hat \mu (k+1)$
and base $PRESS(b) $ or  $PRESAR(b)$ on this approximation.
\label{NSTS.re.bandwidthCV}
\end{Remark}

\subsection{Model-based prediction intervals}
To go from point prediction to prediction intervals, some form of   
resampling is required. Since   model   \eqref{NSTS.eq.model hetero} is driven by 
the stationary sequence $\{W_t\}$, a model-based bootstrap can then be
concocted in which   $\{W_t\}$ is resampled, giving rise
to the bootstrap pseudo-series $\{W_t^*\}$, which in turn gives rise
to   bootstrap pseudo-data $\{Y_t^*\}$ via a fitted
version of model~\eqref{NSTS.eq.model hetero}. To generate a stationary bootstrap
pseudo-series $\{W_t^*\}$, two popular time series resampling methods
are (a) the   stationary bootstrap of 
\tr{\cite{politis1994stationary}}
and 
(b) the AR bootstrap which entails treating the $V_t$ appearing in
eq.~\eqref{NSTS.eq.AR} 
as if they were i.i.d.,   performing an i.i.d.~bootstrap on them, and then
generating $\{W_t^*\}$ via the recursion~\eqref{NSTS.eq.AR} 
driven by the bootstrapped innovations. 
We will use the latter in the sequel because 
it  ties in well with the AR-type predictor of $W_{n+1}$ developed at the end of Section \ref{NSTS.sec.TOPP}, and it is more amenable to the
construction of prediction intervals
\tr {as discussed in \cite{pan2016bootstrap}}.
 In addition,  
\tr{ \cite{kreiss2011range}}
have recently shown that the AR bootstrap---also known as AR-sieve bootstrap
since $p$ is allowed to grow with $n$---can be valid under some 
conditions even if the $V_t$ of eq.~\eqref{NSTS.eq.AR} are not trully i.i.d.

We will now develop an  algorithm for the construction
of model-based prediction intervals;
this is a `forward'  bootstrap algorithm in the terminology 
\tb{of} \tr {\cite{pan2016bootstrap}}
although a `backward'  bootstrap algorithm can also be concocted. 
To describe it in general, let $\check \mu (\cdot) $ and $\check \sigma (\cdot) $
be our chosen estimates of $  \mu (\cdot) $ and $  \sigma (\cdot) $
according to one of 
the abovementioned four methods (NW vs.~LL, regular vs.~predictive);
also let $\check W_t$ denote the resulting proxies for the unobserved 
$W_t$ for $t=1,\ldots,n$. 
Hence, our approximation to the $L_2$--optimal point predictor of 
$ Y_{n+1} $ is 
\begin{equation}
\Pi= \check \mu (n+1) +\check \sigma (n+1)
\left[ \hat \phi_1 \check W_{ n} + \cdots + \hat \phi_p \check W_{n-p+1 } \right] 
\label{NSTS.eq.predictorPi}
\end{equation}
where     
$\hat \phi_1 , \ldots , \hat \phi_p$ are the Yule-Walker estimators of
$  \phi_1 , \ldots ,   \phi_p$ appearing in eq.~\eqref{NSTS.eq.AR}.

As discussed in 
\tb{Chapter 2 of} \tr {\cite{politis2015model}}
the construction of  prediction intervals
will be based on approximating the distribution of the 
  {\it predictive root}: $
 Y_{n+1}  - \Pi $ by that of 
  the   bootstrap predictive root: $
 Y_{n+1}^*  -  \Pi^* $ where the quantities
$Y_{n+1}^*  $ and $  \Pi^* $   are formally  defined 
 in the Model-based (MB) bootstrap algorithm  
outlined below.


\vskip .17in
\begin{Algorithm} {\sc 
Model-based bootstrap for    prediction intervals for
  $ Y_{n+1} $}
\label{NSTS.AlgorithmMB}
\begin{enumerate}
\item Based on the data $Y_1,\ldots, Y_n$, calculate the estimators $\check \mu (\cdot) $ and $\check \sigma (\cdot) $,
and the `residuals'  $\check W_1, \ldots, \check W_n$ using
model   \eqref{NSTS.eq.model hetero}.

\item Fit the AR($p$) model~\eqref{NSTS.eq.AR} to the series $\check W_1, \ldots, \check W_n$ (with $p$ selected by AIC minimization), and obtain the  Yule-Walker estimators $\hat \phi_1 , \ldots , \hat \phi_p$, and the error proxies 
\tr{
$$\check V_t=\check W_{t} -
  \hat \phi_1 \check W_{t-1} - \cdots -  \hat \phi_p \check W_{t-p  }
\ \ \mbox{for} \ \  
\ \  t=p+b+1,\ldots, n.$$ 
}
\tr {Here $b$ is the bandwidth determined by the cross-validation procedure} \tor{of Remark 2.3}.
  
\item 
\begin{enumerate}
\item  Let $\check V_{t}^*$ for $t=  1,\ldots, n,n+1$ be drawn randomly with
replacement from the set  $\{\check {\check {V_t}}$ for \tr {$  t=p+b+1,\ldots, n\} $} 
where
\tr {$\check {\check {V_t}} =\check V_{t}  - (n-p-b)^{-1}\sum_{i=p+b+1}^n  \check V_i$.}
Let $I$ be a random variable drawn from a discrete uniform distribution
on the values  
\tr{$\{  p+b,p+b+1,\ldots, n\} $,} 
and define the bootstrap initial conditions 
$ \check W_t^* =\check W_{t+I}$ for $t= -p+1,\ldots,0$.
Then, create the bootstrap data $\check W_1^*,\ldots,\check W_n^* $ via the  AR recursion 
 $$\check W_t^*= \hat \phi_1 \check W_{t-1}^* + \cdots +  \hat \phi_p \check W_{t-p  }^* + \check V_{t}^*\ \ \mbox{for} \ \  t= 1,\ldots, n. $$ 

\item Create the bootstrap pseudo-series
$Y_1^*,\ldots, Y_n^*$ by the formula
$$ Y_t^*= \check \mu (t) +\check \sigma (t) \check W_t^*
\ \ \mbox{for} \ \  t= 1,\ldots, n. $$ 

\item Re-calculate the estimators $\check \mu^* (\cdot) $ and $\check \sigma^* (\cdot) $  from the bootstrap data
$Y_1^*,\ldots, Y_n^*$. This   gives rises to new
bootstrap `residuals' 
\footnote{\tr {The bootstrap estimators $\check \mu^* (\cdot) $ and $\check \sigma^* (\cdot) $
are based on bandwidth $b'$ determined by Algorithm \ref{doublebootmb.Algorithm} \tnr{given in Appendix \ref{Appendix_Boot}}.
This may be different
from the bandwidth $b$ found using model-based cross-validation.}}
on which an AR($p$) model is again fitted yielding the
bootstrap  Yule-Walker estimators $\hat \phi_1^* , \ldots , \hat \phi_p^*$.

\item Calculate the bootstrap predictor
 $$\Pi^*= \check \mu ^* (n+1) +\check \sigma^* (n+1)
\left[\hat \phi_1^* \check W_{ n} + \ldots + \hat \phi_p^*\check W_{n-p+1 }\right].$$
[Note that in calculating the bootstrap conditional expectation  
of $\check W_{n+1}^* $ given its $p$--past, we have re-defined 
the values $(\check W_{ n}^* , \ldots ,\check W_{n-p+1 }^*)$
to make them match the original $(\check W_{ n} , \ldots ,\check W_{n-p+1 })$;
this is an important part of the `forward' bootstrap procedure for prediction
intervals as discussed in 
\tr {\cite{pan2016bootstrap}}].

\item Calculate a  bootstrap future value 
\tr {$$Y_{n+1}^* = \check \mu  (n+1) +\check \sigma (n+1) \check W_{n+1 }^*$$}
where again 
$\check W_{n+1 }^*=\hat \phi_1 \check W_{n}  + \cdots +  \hat \phi_p \check W_{n-p+1  } + \check V_{n+1}^*$ uses the original values
$(\check W_{ n} , \ldots ,\check W_{n-p+1 })$; recall that $\check V_{n+1}^*$ has  already been generated in step (a) above.
 
\item Calculate the bootstrap root replicate  $Y_{n+1}^*-\Pi^*$.

\end{enumerate}

\item Steps (a)---(f) in the above are repeated  a large number
of   times (say $B$ times),
  and the $B$ bootstrap root replicates 
are collected in the form of an empirical distribution whose  
$\alpha$--quantile is denoted by $q_{ }(\alpha)$.
 
\item
Finally, a $(1-\alpha)$100\% {\it equal-tailed}
prediction interval 
 for $ Y_{n+1} $ is given by
\begin{equation}
[\Pi +q_{ }(\alpha/2), \ 
\Pi + q_{ }(1-\alpha/2)] .
\label{NSTS.eq.predint2.4root}
\end{equation} 
\end{enumerate}
\end{Algorithm}
\vskip .173in
\noindent
It is easy to see that prediction interval~\eqref{NSTS.eq.predint2.4root}
is asymptotically valid (conditionally on $Y_1,\ldots,Y_n$) provided:
(i)  estimators $\check \mu  (n+1)$ and $\check \sigma  (n+1)$ are
consistent for their respective targets $
\mu_{_{[0,1]}} (1) $ and $   \sigma_{_{[0,1]}} (1)$, and
(ii) the AR($p$) approximation is consistent allowing for
the possibility that $p$   grows as $n\to \infty$. 
If  $\check \mu (\cdot) $ and $\check \sigma (\cdot) $
correspond to one of 
the above mentioned four methods (NW vs.~LL, regular vs.~predictive), 
then provision (i) is satisfied under standard conditions 
including the bandwidth condition \eqref{NSTS.eq.new bandwidth}. 
Provision (ii) is also easy to satisfy as long as the spectral density of
the series $\{W_t\}$ is continuous and bounded away from zero; see e.g. 
Lemma 2.2 of 
\tr{\cite{kreiss2011range}}.

Although desirable, asymptotic validity does not tell the whole story. 
A prediction interval can be thought to be successful if it also manages
to capture the finite-sample variability of the estimated quantities
such as $\check \mu (\cdot) $, $\check \sigma (\cdot) $
and $\hat \phi_1, \hat \phi_2, \ldots$. Since this finite-sample variability
vanishes asymptotically, the performance of a
prediction interval such as~\eqref{NSTS.eq.predint2.4root}
must be gauged by finite-sample simulations. \tr {Results of these simulations
are shown in Section \ref{NSTS.Numerical}.}

\section{Model-free inference} 
\label{NSTS.Model-free inference}

Model (\ref{NSTS.eq.model hetero}) is a flexible way to account for a
time-changing  mean and variance of $Y_t$. 
However, nothing precludes that the time series $\{Y_t $ for $t\in {\bf Z}   \}$ has a nonstationarity
in its third (or higher moment), and/or in some other feature of its $m$th
marginal distribution.
A way to address this difficulty, and at the same time give a fresh perspective to the problem, is provided by the 
Model-Free Prediction Principle of  Politis (2013, 2015).

The key towards Model-free inference is to be able to construct an invertible transformation 
$H_n: \underline{Y}_n \mapsto \underline \epsilon_{n}$
where $\underline \epsilon_{n}=(\epsilon_{1}, \ldots, \epsilon_{n} )'$
is  a random vector with i.i.d.~components. In order to do this in our context, 
\tb {let some $m\geq1$, and denote by ${\cal L}(Y_{t },Y_{t-1 },\ldots,Y_{t-m+1}) $ 
 the $m$th marginal of the time series { $Y_{t}$ }, i.e. the joint probability law of the vector
$(Y_{t },Y_{t-1 },\ldots,Y_{t-m+1})'$. Although we  abandon  model~(\ref{NSTS.eq.model hetero}) 
in what follows, we still want to employ nonparametric smoothing for estimation;  thus,
we must assume that ${\cal L}(Y_{t },Y_{t-1 },\ldots,Y_{t-m+1})$ \tor{changes}  smoothly (and slowly) with $t$.}

\tb {
\begin{Remark} [Quantifying smoothness--model-free case] \rm
As in Remark \ref{NSTS.re.smooth}, we can formally quantify 
smoothness by mapping the index set $\{1, \ldots, n\}$ onto the interval $[0,1]$. 
Let $\underline{s}=(s_0,s_1,\ldots, s_{m-1})'$, and
 define the distribution function of the $m$th
marginal by 
 $$D_{t }^{(m)}(\underline{s})=P\{ Y_t\leq s_0,
 Y_{t-1}\leq s_1,\ldots,Y_{t-m+1}\leq s_{m-1}  \}. $$ 
Let $a_t=(t-1)/n$   as before, and   assume that we can write 
\begin{equation}
D_{t }^{(m)}(\underline{s})=  D^{^{[0,1]}}_{a_t}(\underline{s})
\ \ \mbox{for} \ t=1, \ldots, n.
 \label{NSTS1.eq.smooth2}
\end{equation}
We can now quantify smoothness by assuming
 that, for each fixed $\underline{s}$, 
the function $D^{^{[0,1]}}_{x}(\underline{s})$ is continuous and
  smooth in $x\in [0,1]$, i.e.,   possesses $k$ continuous derivatives.
As in Remark~\ref{NSTS.re.smooth}, here as well it seems to be sufficient
that $D^{^{[0,1]}}_{x}(\underline{s})$ is   continuous in $x$ but only piecewise smooth. 
\label{NSTS.re.smooth2} 
\end{Remark}
}

\tb {
A convenient way to ensure both the smoothness and data-based consistent estimation of
${\cal L}(Y_{t },Y_{t-1 },\ldots,Y_{t-m+1})$ is to assume that, \tor{for all $t$,}}

\begin{equation}
 Y_{t } =
{\bf f}_{t}(W_{t },W_{t-1 },\ldots,W_{t-m+1})
\label{NSTS1.eq.book.9.24}
\end{equation}

\noindent
\tb {for some function ${\bf f}_{t}$($w$) that is smooth in both arguments $t$ and $w$, and some strictly stationary
and weakly dependent,  univariate time series ${W_t}$; without loss of
generality, we may assume that ${W_t}$ is a Gaussian time series. 
In fact, Eq. (\ref{NSTS1.eq.book.9.24}) with ${\bf f}_{t}$($\cdot$) not depending on
$t$ is a familiar assumption in studying non-Gaussian and/or long-range dependent 
stationary processes---see e.g. \cite{samorodnitsky1994stable}. 
By allowing ${\bf f}_{t}$($\cdot$)  to vary smoothly (and slowly) with $t$,
 Eq. (\ref{NSTS1.eq.book.9.24})  can be used to describe a rather general
class of locally stationary processes.
Note that model~(\ref{NSTS.eq.model hetero}) is a special case
of Eq. (\ref{NSTS1.eq.book.9.24}) with $m=1$, and
 the function ${\bf f}_{t}$($w$) being affine/linear in $w$.
Thus, for concreteness and easy comparison with the model-based case of
Eq. (\ref{NSTS.eq.model hetero}), we will focus in the sequel on the case $m=1$. Section \ref{NSTS.sec.higher-dimensional marginals}
discusses how to handle the case $m > 1$.
}

\subsection{Constructing the theoretical transformation}
\label{NSTS.sec.CTT}

\tb {Hereafter, adopt the setup of \tor{ Eq. (\ref{NSTS1.eq.book.9.24})} with $m=1$,}
and let 
$$D_{t }(y)=P\{ Y_t\leq y  \}$$ denote  the 1st marginal distribution of time series $\{Y_t \}$.
Throughout Section \ref{NSTS.Model-free inference},
the default assumption will be that $D_{t }(y)$ is (absolutely) continuous in $y$ for all $t$; however, a departure from this assumption will be discussed 
in Section \ref{NSTS.discrete data}.  

We now define new variables via the probability integral transform, i.e., let 
\begin{equation}
U_t =  D_t(Y_t) \ \ \mbox{for} \ t=1,\ldots,n;
 \label{NSTS1_unif.eq.modelT}
\end{equation}
the assumed continuity of $D_{t }(y)$  in $y$  implies that 
  $U_1,\ldots, U_n$ are random variables having   distribution Uniform $ (0,1)$.
However, $U_1,\ldots, U_n$ are dependent; to transform them to 
independence, a preliminary transformation towards Gaussianity is helpful as
 discussed in 
\tr{\cite{Politis2013}}.
Letting $\Phi$ denote the cumulative distribution function (cdf) of the standard normal distribution, we define
\begin{equation}
Z_t = \Phi^{-1} (U_t)  \ \ \mbox{for} \ t=1,\ldots,n;
 \label{NSTS1_norm.eq.modelT}
\end{equation}
it then follows that $Z_1,\ldots, Z_n$ are standard normal---albeit 
correlated---random variables. 

Let $ \Gamma_n $ denote  the $n\times n$ covariance matrix
of the   random vector $\underline{Z}_n=(Z_1,\ldots, Z_n)' $.
Under standard   assumptions,
e.g. that the spectral density of the series $\{Z_t\}$
is continuous and bounded away from zero,\footnote{If the spectral density 
is equal to zero over an interval---however small---then the time series $\{Z_t\}$ is perfectly predictable 
based on its infinite past, and the same would be true for 
 the time series $\{Y_t\}$; see Brockwell and Davis (1991, Theorem 5.8.1)
on Kolmogorov's formula.}   the matrix $ \Gamma_n $ is invertible 
when  $n$ is large enough. Consider the 
Cholesky decomposition   $ \Gamma_n  =  C_n   C_n'$ 
where $  C_n$ is (lower) triangular, and  construct the 
{\it whitening} transformation:
\begin{equation}
\label{NSTS.eq.whitenfilterT}
 \underline \epsilon_{n}=   C_n^{-1} \underline{Z}_n .
\end{equation}
It then follows that the entries of $\underline \epsilon_{n}=(\epsilon_1, \ldots, \epsilon_n)'$ are uncorrelated~standard normal.
Assuming that the random variables $Z_1,\ldots, Z_n$ were {\it jointly} normal, 
this can be strenghtened to claim that $
\epsilon_1, \ldots, \epsilon_n$ are i.i.d.~$N(0,1)$; see 
Section \ref{NSTS.sec.higher-dimensional marginals} for further discussion.
Consequently, the transformation 
of the dataset $\underline{Y}_n=(Y_1,\ldots, Y_n)' $
to the vector $\underline \epsilon_{n}$ with i.i.d.~components has been
achieved as required in premise (a) of the 
Model-free Prediction Principle. 
 Note that all the steps in the transformation, i.e., 
   eqs.~(\ref{NSTS1_unif.eq.modelT}), (\ref{NSTS1_norm.eq.modelT}) and (\ref{NSTS.eq.whitenfilterT}), are invertible; hence, the composite
transformation $H_n: \underline{Y}_n \mapsto \underline \epsilon_{n}$
is invertible as well.

\subsection{Kernel estimation of the `uniformizing' transformation}
\label{NSTS.sec.KEUT}

We first focus on estimating the `uniformizing' part of the transformation,
i.e., eq.~(\ref{NSTS1_unif.eq.modelT}).
Recall that the Model-free setup implies that the function $D_t(\cdot)$ changes smoothly (and slowly) with~$t$; hence, local constant   and/or local linear fitting can be used to estimate it. 
Using local constant, i.e.,  kernel estimation, a consistent  estimator of the marginal distribution 
$D_{t }(y)$ is given by:
\begin{equation}
\hat D_{t }(y) = \sum_{i=1}^{T} {\bf 1}\{ Y_{t_{i}}\leq y\} 
 \tilde K (\frac{t -t_{i}}{b})
\label{NSTS.eq.hatD}
\end{equation}
where $\tilde K (\frac{t -t_{i}}{b}) =  K (\frac{t -t_{i}}{b})/
  \sum_{j=1}^{T}K (\frac{ t-t_{j}}{b}) $. 
Note that the kernel estimator \eqref{NSTS.eq.hatD} is {\it one-sided}
for the same reasons discussed in Remark \ref{NSTS.re.onesided}.
Since $\hat D_{t }(y)$ is a step function in $y$, a smooth estimator 
can be defined as:
\begin{equation}
\bar D_{t }(y) = \sum_{i=1}^{T} \Lambda(\frac {y-Y_{t_{i}}} {{h}_0})\tilde K (\frac{t-t_{i}}{b})
\label{NSTS.eq.barD}
\end{equation}
where ${h}_0$ is a secondary bandwidth.
 Furthermore, as in Section \ref{NSTS.sec.trend}, we can let $T=t$ or $T=t-1$
leading to a {\bf fitted vs.~predictive} way to estimate $D_{t }(y)$
by either $\hat D_{t }(y)$  or $\bar D_{t }(y)$. 
\tr {Cross-validation is used to determine the bandwidths $h_0$ and $b$ \tor{; details} 
are described in Section \ref{NSTS.sec.MF.cv}.}


\subsection{Local linear estimation of the `uniformizing' transformation}
\label{NSTS.sec.LLEUT}

Note that the kernel estimator 
$\hat D_{t }(y)$ defined in eq.~\eqref{NSTS.eq.hatD}
is just the Nadaraya-Watson smoother, i.e., local average,  of the variables 
$u_1,\ldots, u_n$ where $u_i = {\bf 1}\{ Y_i\leq y\}$.
 Similarly, $\bar D_{t }(y)$ defined in eq.~\eqref{NSTS.eq.barD}
is just the Nadaraya-Watson smoother of the variables 
$v_1,\ldots, v_n$ where $v_i =\Lambda(\frac {y-Y_i} {{h}_0})$. 
In either case, it is only natural to try to consider a local
linear smoother as an alternative to Nadaraya-Watson especially
since, once again, our interest lies on the boundary, i.e., the
case $t=n$. 
  
Let 
\tr {$\hat D_{t }^{LL}(y)$ and $\bar D_{t}^{LL}(y)$} 
denote the local linear  estimators of $D_{t }(y)$
based on either the indicator variables ${\bf 1}\{ Y_i\leq y\}$
or the smoothed variables $\Lambda(\frac {y-Y_i} {{h}_0})$ 
\tr {respectively}. 
Keeping $y$ fixed, 
\tr {$\hat D_{t }^{LL}(y)$ and $\bar D_{t}^{LL}(y)$}
\tr {exhibit} good behavior \tr{for estimation at the boundary,} e.g.
smaller bias than either $\hat D_{t }(y)$ \tr{and} $\bar D_{t }(y)$
\tr{respectively}.
However, there is no guarantee that
\tr {these will be}
proper distribution \tr{functions} as a function of $y$, 
i.e., being nondecreasing in $y$ with a left limit of 0 and
a right limit of 1; see 
\tr{\cite{li2007nonparametric}}
for a discussion.

There have been several proposals in the literature to address this issue.
An interesting one   is the adjusted Nadaraya-Watson estimator 
of 
\tr {\cite{Hall1999}}
which, however, is   tailored
towards nonparametric autoregression estimation rather than our 
setting where $Y_t$ is regressed on $t$. 
Coupled with the fact that we are interested in the 
boundary case $t=n$, the equation yielding the adjusted Nadaraya-Watson  
  weights   \tb{do not always} admit a solution. 

\tr {One proposed solution put forward by
\cite{hansen2004nonparametric}} 
 \tr {involves}
 a straightforward adjustment to 
the local linear estimator of a conditional distribution function 
that maintains its favorable asymptotic properties.
The local linear  versions of $\hat D_t(y) $ and $\bar D_t(y) $
adjusted via 
  Hansen's (2004) 
proposal are
  given as follows:
\begin{equation}
\label{NSTS.eq.sll_cdf}
\hat D_t^{LLH}(y) = \frac{\sum_{i=1}^{T} w_{i}^\diamond {\bf 1}(Y_{i} \le y)}{\sum_{i=1}^{T} w_{i}^\diamond}  
\ \ \mbox{and} \ \ 
\bar D_t^{LLH}(y) = \frac{\sum_{i=1}^{T} w_{i}^\diamond 
\Lambda(\frac {y-Y_i} {{h}_0})}{\sum_{i=1}^{T} w_{i}^\diamond} .
\end{equation}
The weights $w_{i}^\diamond $ are defined by
\begin{align}
\label{NSTS.eq.sll_diamond}
w_{i}^\diamond = \begin{cases}
                 \ 0  & \ \mbox{when} \ \hat \beta(t-t_{i}) > 1 \\
                 \ w_{i}(1 - \hat \beta(t-t_{i})) & \ \mbox{when} \ \hat \beta(t-t_{i}) \le 1
                 \end{cases}
\end{align}
where 
\begin{equation}
\label{NSTS.eq.sll_weights}
w_{i} = \frac{1}{b}\ K(\frac{t-t_{i}}{b})
\ \ \mbox{and} \ \ 
\hat \beta =\frac{\sum_{i=1}^T\ w_{i}(t-t_{i})}{ \sum_{i=1}^T\ w_{i}(t-t_{i})^2 }.
\end{equation}

\tr 
{As with eq.~\eqref{NSTS.eq.hatD}and~\eqref{NSTS.eq.barD}, we can let $T=t$ or $T=t-1$ in the above, 
leading to a {\bf fitted vs.~predictive} 
local linear estimators of  $D_{t }(y)$,
by either $\hat D_t^{LLH}(y)$ or $\bar D_t^{LLH}(y)$}.

\tr{
\subsection{Uniformization using Monotone Local Linear Distribution Estimation}
}
\label{NSTS.sec.MLLDE}

\tr
{Hansen's \tb {(2004)} proposal replaces  negative weights by zeros, and then renormalizes the nonzero
weights. The problem here is that if estimation is performed on the boundary (as in the case
with \tor{one-step ahead prediction of} time-series), negative weights are crucially needed in order to ensure \tor{the}
extrapolation takes place with minimal bias.} 
\tc{A recent proposal by  \cite{das2017nonparametric}
addresses this issue by modifying the original, possibly nonmonotonic local linear distribution estimator $\bar D_{t}^{LL }(y) $ to construct a monotonic version 
denoted by $\bar {D}_{t}^{LLM }(y)$.}

\medskip
\tc{The Monotone Local Linear Distribution Estimator $\bar {D}_{t}^{LLM }(y)$ can be   constructed by Algorithm \ref{Monotone_Density_Algo} given below.}

\medskip

%
%

\begin{Algorithm}{
\bf{Monotone Local Linear Distribution Estimation}
\label{Monotone_Density_Algo}
}
\label{Monotone_density_LL}
\begin{enumerate}
\item Recall that the derivative of $\bar {D}_{t}^{LL  }(y) $ 
with respect to $y$ is given by 
$$  \bar d_{t}^{LL}(y)=\frac{\frac {1} {{h}_0} \sum_{j=1}^{n } w_j  \lambda(\frac {y-Y_{ {j}}} {{h}_0})  }{\sum_{j=1}^{n } w_j  }  
 $$
where $\lambda(y)$ is the derivative of  $\Lambda(y)$.
\item Define a nonnegative version of $  \bar d_{t}^{LL}(y)$ as
$  \bar d_{t}^{LL+}(y)=\max (\bar d_{t}^{LL}(y), 0)$.
\item To make the above a proper density function, renormalize
it to area one, i.e., let 
\begin{equation}
\label{eq.densityMLL}
\bar d_{t}^{LLM }(y) = \frac{\bar d_{t}^{LL+}(y)  } {\int_{-\infty}^\infty \bar d_{t}^{LL+}(s)ds  }.
\end{equation}
\item Finally, define $\bar {D}_{t}^{LLM }(y) =\int_{-\infty}^y \bar d_{t}^{LLM }(s) ds.$
\end{enumerate}
\end{Algorithm}

 The above modification of the local linear estimator
allows one to maintain monotonicity while retaining the negative weights that
are helpful in problems which involve estimation at the boundary.
{As with eq.~\eqref{NSTS.eq.hatD}and~\eqref{NSTS.eq.barD}, we can let $T=t$ or $T=t-1$ in the above, 
leading to a {\bf fitted vs.~predictive} 
local linear estimators of  $D_{t }(y)$ that are monotone.
}

\tbl{Different algorithms could also be employed for performing monotonicity correction on the original estimator $\bar {D}_{t}^{LL  }(y) $; these are discussed in detail in  \cite{das2017nonparametric}.  In practice, Algorithm \ref{Monotone_Density_Algo} is
preferable because it is the fastest in term of implementation; notably, 
 density estimates can be obtained in a fast way (using the Fast Fourier Transform) 
using standard functions in statistical software such as R. 
Computational speed is particularly important in constructing
bootstrap prediction intervals
 since a large number of estimates of $\bar D_{t }^{LLM}(y)$ must be computed;
the same is true for cross-validation implementation which is addressed next.}
\tr
{
\subsection{Cross-validation Bandwidth Choice for Model-Free Inference}
\label{NSTS.sec.MF.cv}
}

\tr {
There are two bandwidths, $b$ and $h_0$, required to construct the estimators $\bar {D}_{t}(y)$,
$\bar {D}_{t}^{LLH}(y)$ and $\bar {D}_{t}^{LLM}(y)$. This discussion first focuses on choice of $b$
as it is \tb{the most} crucial of the two. The following steps are recommended:}
\newpage

\begin{Algorithm} {
BANDWIDTH DETERMINATION FOR MODEL-FREE INFERENCE
\label{Bandwidth_MF_Algo}
}
\begin{enumerate}

\item 
\tr{
Perform the uniformizing transform described in \eqref{NSTS1_unif.eq.modelT} over the given time-series
dataset $Y_1,\ldots, Y_n$ using either of the estimators $\bar {D}_{t}(y)$, $\bar {D}_{t}^{LLH}(y)$ or
$\bar {D}_{t}^{LLM}(y)$ over \tor{$q$} pre-defined bandwidths \tb{that span an interval of possible values}.
}
\item
\tr{
Calculate the value of the Kolmogorov-Smirnov (KS) test statistic using the uniform distribution $U[0,1]$
as reference for each of these $q$ cases.
}
\item
\tr{
\tb{From the full list of \tor{$q$} values given in step (1) above} pick \tb {a} pre-defined number of bandwidths, \tb{say this is $p$,} whose \tb{corresponding} KS test statistic values are minimum. 
These represent the bandwidths which achieved the best transformation to
`uniformity' using $\bar {D}_{t}(y)$, $\bar {D}_{t}^{LLH}(y)$ or $\bar {D}_{t}^{LLM}(y)$.
}

\item

\tr
{
Obtain the best bandwidth $b$ among these $p$ values by using one-sided cross-validation in a similar manner
as described for the Model-Based case in Section \ref{NSTS.sec.trend}. For this purpose let  $k<n$, and suppose
only subseries $Y_1,\ldots, Y_k$ has been observed. Denote $\hat Y_{k+1}$ the best predictor of $Y_{k+1}$
based on the data $Y_1,\ldots, Y_k$  constructed using $\bar {D}_{t}(y)$, $\bar {D}_{t}^{LLH}(y)$ or $\bar {D}_{t}^{LLM}(y)$
and a value of $b$ selected among the $p$ values obtained above.
Since $Y_{k+1}$ is known, the quality of the predictor can be assessed. So, for each value of $b$
we can form either $PRESS(b)=\sum_{k=k_o}^{n-1} ( \hat Y_{k+1} - Y_{k+1})^2 $
or  $PRESAR(b)=\sum_{k=k_o}^{n-1} | \hat Y_{k+1} - Y_{k+1}| $; here
$k_o$ should be big enough so that estimation is accurate, e.g., $k_o$ can be of the order of $\sqrt{n}$.
We then select the  bandwidth   $b$ that
minimizes $PRESS(b)$; alternatively, we can choose to minimize $PRESAR(b)$ if
 an   $L_1$ measure of loss is preferred.
 }
 
 \item
 \tr
 {
 Coming back to the problem of selecting $h_0$,  as
 in  \cite{Politis2013},  our final choice is  $h_0=h^2$ where   $h=b/n$.
Note that an initial choice of $h_0$ needed
(\tb{to perform} uniformization, KS statistic generation and cross-validation to determine the optimal bandwidth $b$) 
can be set by any  plug-in rule; the  effect of choosing an initial value of $h_0$ 
 is minimal.
}

\end{enumerate}
\end{Algorithm}

\tb{The above algorithm needs large data sizes in order to work well. In the case of smaller data sizes of, say,   \tbl{a hundred or so} data points,
 it is recommended to omit steps (1)--(3) and
directly perform steps (4) and (5) using the full range of \tor{$q$} pre-defined bandwidths. 
}

\subsection{Estimation of the whitening transformation} 
\label{NSTS.seq.whiteningtransformation}

To implement the whitening transformation \eqref{NSTS.eq.whitenfilterT},
it is necessary to estimate $ \Gamma_n $, i.e., the $n\times n$ covariance matrix
of the   random vector $\underline{Z}_n=(Z_1,\ldots, Z_n)' $
where the $Z_t$ are the normal random variables
 defined in eq.~\eqref{NSTS1_norm.eq.modelT}.

As discussed in the analogous \tr {model-based} problem in Section \ref{NSTS.sec.TOPP}, there
are two approaches towards positive definite estimation of $ \Gamma_n $ based on the sample  $Z_1,\ldots, Z_n$. They are both based on the sample autocovariance 
defined as $\breve \gamma_k = n^{-1}\sum_{t=1}^{n-|k|}Z_tZ_{t+|k|}$
for $|k|<n$; for $|k|\geq n$, we define $\breve \gamma_k =0$.

\begin{itemize}
\item [A.] Fit a causal AR($p$) model to the data $Z_1,\ldots, Z_n$ with $p$
obtained via AIC  minimization. Then, let $\hat \Gamma_n^{AR}$ be the 
$n\times n$ covariance matrix associated with the fitted AR model. 
Let $\hat \gamma_{|i-j|}^{AR}$ denote the $i,j$  element of the
Toeplitz matrix $\hat \Gamma_n^{AR}$. Using the Yule-Walker
equations to fit the AR model implies that 
$\hat \gamma_{k}^{AR}= \breve \gamma_k $ for $k=0,1,\ldots, p.$ 
For $k> p$, $\hat \gamma_{k}^{AR}$ can be found by solving
(or just iterating) the 
difference equation that characterizes the (fitted) AR model;
R automates this process via the {\tt ARMAacf()} function.

\item [B.]  
Let $\hat \Gamma_n = \left[ \hat   \gamma_{|i-j|} \right]_{i,j=1}^n $
be the  matrix estimator of  
\tr{\cite{mcmurry2010banded}}
where $ 
\hat   \gamma_s= \kappa(|s|/l) \breve \gamma_s $.
Here,   $\kappa(\cdot)$ can be any member of the 
 {\it flat-top} family of compactly supported 
functions defined in 
\tr {\cite{politis2001nonparametric}}
the simplest choice---that has been shown to 
work well in practice---is the trapezoidal, 
i.e.., $ \kappa(x) =  (\max \{1, 2 - |x|\} )^+$ 
where $(y)^+=\max \{y,0\}$ is the positive part function, \tor{\cite{politis1994stationary}}.
Our final estimator of $  \Gamma_n$ will be 
$\hat \Gamma_n ^\star$ which is a 
a positive definite version of $\hat \Gamma_n $  that is banded and
Toeplitz; for example,  $\hat \Gamma_n ^\star$ may be obtained by  
 shrinking   $\hat \Gamma_n $  
towards white noise or  towards a second order estimator as described in
\tr {McMurry and Politis (2015).}
 
\end{itemize}

Estimating the `uniformizing' transformation
$D_t(\cdot)$ and  the whitening trasformation  based on $  \Gamma_n$ 
  allows us to estimate the 
transformation $H_n: \underline{Y}_n \mapsto \underline \epsilon_{n}$.
However, in order to put the Model-Free Prediction Principle to work, 
we also need to   estimate the transformation $H_{n+1}$ (and   its inverse).
To do so, we need a positive definite estimator for 
the matrix $  \Gamma_{n+1}$; this can be accomplished by 
either of the two ways discussed in the above.

\begin{itemize}
\item [A$'$.]  
Let $\hat \Gamma_{n+1}^{AR}$ be the 
$(n+1)\times (n+1)$ covariance matrix associated with the fitted AR($p$) model.  

\item [B$'$.]  
Denote by $\hat \gamma_{|i-j|}^\star$ the $i,j$ element
of  $\hat \Gamma_n^\star$ for $i,j=1,\ldots,n$.
Then, define $\hat \Gamma_{n+1} ^\star$  to be the symmetric, banded Toeplitz $(n+1)\times (n+1)$ matrix
with $ij$ element given  by $\hat \gamma_{|i-j|}^\star$   
when $|i-j|<n$. 
Recall that  $\hat \Gamma_n^\star$ is 
 banded \tr {with banding parameter $l$ as discussed in \cite{mcmurry2015high}}, so it is only natural to assign zeros to the 
two  $ij$ elements of $\hat \Gamma_{n+1} ^\star$  that satisfy
$|i-j| = n$, i.e., the bottom left and the top right.
 
\end{itemize} 

   \noindent
Consider the `augmented' vectors
 $\underline{Y}_{n+1}= (Y_1,\ldots,Y_n, Y_{n+1})'$, 
$\underline{Z}_{n+1}=(Z_1,\ldots,Z_n,   Z_{n+1})'$
and $\underline \epsilon_{n+1}=(\epsilon_1,\ldots, \epsilon_n, \epsilon_{n+1})'$
where the values $Y_{n+1}, Z_{n+1}$ and $\epsilon_{n+1}$
are yet unobserved. 
We now show how to obtain 
the inverse transformation $H_{n+1}^{-1}: 
\underline \epsilon_{n+1} \mapsto 
\underline{Y}_{n+1} $. Recall that $\underline \epsilon_{n}$ and $
\underline{Y}_{n } $ are related in a one-to-one way via 
  transformation $H_{n}$, so the values $ Y_1,\ldots,Y_n$
are obtainable by $\underline{Y}_{n } =H_{n}^{-1}(\epsilon_{n}).$
Hence, we just need to show how to create the unobserved
$Y_{n+1}$   from $\underline \epsilon_{n+1} $; this is 
done in the following three steps.

\begin{Algorithm} {
\tb{GENERATION OF UNOBSERVED DATA\tor{POINT} FROM \tor{FUTURE }INNOVATIONS}
\label{Pred_MF_Algo}
}

\begin{itemize}
 
 \smallskip
 \item [i.] Let 
\begin{equation}
\label{NSTS.eq.CholPred}
\underline{Z}_{n+1}=  C_{n+1} \underline \epsilon_{n+1}
\end{equation}
 where $  C_{n+1}$ is the (lower) triangular Cholesky factor of
(our positive definite estimate of)  $  \Gamma_{n+1} $.
From the above, it follows that 
 
\begin{equation}
\label{NSTS.eq.invtrans1}
{Z}_{n+1}=\underline c_{n+1} \underline \epsilon_{n+1}
\end{equation}
where $\underline c_{n+1}=(c_1,\ldots,c_n,c_{n+1}) $ is 
a row vector consisting of  the last row of
matrix $  C_{n+1}$.

\item [ii.] Create the uniform random variable 
\begin{equation}
\label{NSTS.eq.invtrans2}
{U}_{n+1}= \Phi(Z_{n+1}) . 
\end{equation}

\item [iii.] 
Finally, define  
\begin{equation}
\label{NSTS.eq.invtrans3}
Y_{n+1} =   D_{n+1}^{-1} ({U}_{n+1}); 
\end{equation}
of course, in practice, the above will be based on an
estimate of  $ D_{n+1}^{-1}(\cdot)$. 
\end{itemize}
\end{Algorithm}

\noindent
Since $\underline{Y}_{n } $  has already been created using (the
first $n$ coordinates of) $\underline \epsilon_{n+1}$, the above completes
the construction of $\underline{Y}_{n+1} $ based on $\underline \epsilon_{n+1}$,
i.e., the mapping  $H_{n+1}^{-1}: 
\underline \epsilon_{n+1} \mapsto \underline{Y}_{n+1} $. 

\subsection{Model-free predictors and   prediction intervals}
\label{NSTS.sec.prediction intervals}

In the previous sections, it was shown how the construct the  transformation 
$H_n: \underline{Y}_n \mapsto \underline \epsilon_{n}$  and its
inverse $H_{n+1}^{-1}: 
\underline \epsilon_{n+1} \mapsto \underline{Y}_{n+1} $,
where the random variables $\epsilon_{1},  \epsilon_2,  \ldots,$
are i.i.d. Note that by combining eq.~(\ref{NSTS.eq.invtrans1}), (\ref{NSTS.eq.invtrans2}) and (\ref{NSTS.eq.invtrans3}) we can write the formula: 
$$
  Y_{n+1} =   D_{n+1}^{-1}\left( \Phi( \ \underline c_{n+1} \underline \epsilon_{n+1} )\right).
$$
Recall that $\underline c_{n+1} \underline \epsilon_{n+1}
= \sum_{i=1}^n c_i \epsilon_i+c_{n+1} \epsilon_{n+1} $;
hence, the above can be compactly denoted as 
\begin{equation}
\label{NSTS.eq.pred.equation}
  Y_{n+1} =  g_{n+1}(\epsilon_{n+1}) 
\ \ \mbox{where} \ \ 
g_{n+1}(x)=
 D_{n+1}^{-1}\left( \Phi \left( \  \sum_{i=1}^n c_i \epsilon_i+c_{n+1}x \right) \right).
\end{equation}
Eq.~\eqref{NSTS.eq.pred.equation} is the predictive equation required in the 
Model-free Prediction Principle; 
conditionally on $\underline{Y}_{n } $, it can be used like a model
equation in computing the $L_2$-- and $L_1$--optimal point predictors of 
 $Y_{n+1} $. We will give these in detail as part of the general
algorithms for the construction of Model-free predictors and prediction intervals.

\begin{Algorithm} 
\label{NSTS.Algo.BasicMF}
{\sc   Model-free (MF) predictors and prediction intervals for $Y_{n+1} $}
 
\begin{enumerate}
\item Construct $U_1,\ldots, U_n$ by eq.~(\ref{NSTS1_unif.eq.modelT}) with 
$D_t(\cdot)$ estimated by either $\bar D_t(\cdot)$ \tr {, $\bar D_t^{LLH}(\cdot)$
or $\bar D_t^{LLM}(\cdot)$};
for \tr {all the 3 types of estimators}, use the respective formulas with $T=t$.
\item Construct $Z_1,\ldots, Z_n$ by eq.~(\ref{NSTS1_norm.eq.modelT}), and use the
methods of Section \ref{NSTS.seq.whiteningtransformation} to estimate
$\Gamma_n$ by either $\hat \Gamma_n^{AR}$ or $\hat \Gamma_n^\star$.
\item  
Construct $\epsilon_1,\ldots, \epsilon_n$ by eq.~(\ref{NSTS.eq.whitenfilterT}),
and let $\hat F_n$ denote their empirical distribution.

\item 
The Model-free $L_2$--optimal point predictor  of 
 $Y_{n+1} $ is then \\
\tr{$$ \hat Y_{n+1}= \int  g_{n+1}(x ) dF_n(x)
= \frac{1}{n} \sum_{i=1}^n g_{n+1}(\epsilon_i ) $$}
where the function $g_{n+1}$ is defined in the 
predictive equation \eqref{NSTS.eq.pred.equation}
with $D_{n+1}  (\cdot)$ being again estimated by either $\bar D_{n+1}  (\cdot)$
\tr{
,
$ \bar D_{n+1}^{LLH}  (\cdot)$
or $ \bar D_{n+1}^{LLM}  (\cdot)$
}  
\tr {all} 
with $T=t$. 

\item 
The Model-free $L_1$--optimal point predictor  of 
 $Y_{n+1} $ is given by the median of the set
$\{  g_{n+1}(\epsilon_i )$ for $i=1,\ldots, n\}$.

\item 
Prediction intervals for $Y_{n+1} $ with prespecified coverage
probability can be constructed via the Model-free Boootstrap of  
Algorithm \ref{MF3short.Algorithm1} 
based on either the $L_2$-- or $L_1$--optimal point predictor.
\end{enumerate}
\end{Algorithm} 
\vskip .13in
\noindent
Algorithm \ref{NSTS.Algo.BasicMF}  used the construction of
$\bar D_t(\cdot)$ \tr {,
$\bar D_t^{LLH}(\cdot)$ 
\tr {or $\bar D_t^{LLM}(\cdot)$ }
}
  with $T=t$;  using $T=t-1$ instead,  
 leads to the {\it predictive} version of the algorithm.
\vskip .173in
 \begin{Algorithm} 
\label{NSTS.Algo.PMF}
{\sc   Predictive Model-free (PMF) predictors and prediction intervals for $Y_{n+1} $} \\
The algorithm is identical to Algorithm \ref{NSTS.Algo.PMF}
except for using $T=t-1$ instead of $T=t $ in the construction
of $\bar D_t(\cdot)$ 
\tr{
,  $\bar D_t^{LLH}(\cdot)$ and $\bar D_t^{LLM}(\cdot)$.
}
 \end{Algorithm} 
\vskip .13in
\begin{Remark} \rm
Under a model-free setup of a locally stationary time series, 
\tr{\cite{paparoditis2002local}}
proposed the  Local Block Bootstrap (LBB)
in order to generate pseudo-series $Y_1^*, \ldots, Y_n^*$
whose probability structure mimics that of the observed data
$Y_1, \ldots, Y_n$. The 
Local Block Bootstrap 
 has been found useful for the construction of 
confidence intervals; see 
\tr {\cite{dowla2003locally} and \cite{dowla2013local}.}
However, it is unclear if/how the LBB can be employed for the
 construction of predictors and 
prediction  intervals for $Y_{n+1}$. 
\end{Remark}

\vskip .1in
Recall that when the theoretical transformation $H_n$
is employed, the 
variables $\epsilon_1,\ldots, \epsilon_n$ are   i.i.d.~$N(0,1)$. 
Due to the fact that features of $H_n$ are unknown and must be
estimated from the data, the practically available 
variables $\epsilon_1,\ldots, \epsilon_n$ are only
approximately i.i.d.~$N(0,1)$.  However,  their empirical distribution of
   $\hat F_n$ converges to $F=\Phi$ as $n\to \infty$. 
Hence, it is possible to use the limit distribution  $F=\Phi$ in 
instead of   $\hat F_n$  in both the construction of point predictors and
the prediction intervals; this is an application of the 
Limit Model-Free (LMF) approach  
\tb{as discussed in \cite{politis2015model}.}
 
The LMF Algorithm is simpler than  Algorithm \ref{NSTS.Algo.PMF}
as the first three steps of the latter  
can be omitted. As a matter of fact, the LMF Algorithm is totally
based on the inverse transformation $H_{n+1}^{-1}: 
\underline \epsilon_{n+1} \mapsto \underline{Y}_{n+1} $; the
forward    transformation 
$H_n: \underline{Y}_n \mapsto \underline \epsilon_{n}$ is not needed
at all. 
But for the inverse transformation it is sufficient to estimate
$D_t(y)$ by the step functions $\hat D_{t}  (y)$ 
\tr{,   $ \hat D_{t}^{LLH}  (y)$ 
or $ \hat D_{t}^{LLM}  (y)$ 
}
with the understanding that 
their inverse must be a {\it quantile} inverse;
recall that the   quantile  inverse of a distribution $D(y)$
is defined as 
$  D ^{-1}(\beta)=\inf \{ y $ such that $ D (y)\geq \beta \}$. 
 
\vskip .17in
\begin{Algorithm} 
\label{NSTS.Algo.LMF}
{\sc   Limit Model-free (LMF) predictors and prediction intervals for $Y_{n+1} $}
 
\begin{enumerate}
 
 \item 
The LMF $L_2$--optimal point predictor  of 
 $Y_{n+1} $ is  
\begin{equation}
\label{NSTS.eq.LMFpp}
\tr{
\hat Y_{n+1}= \int g_{n+1}(x ) d\Phi(x)
}
\end{equation}
where the function $g_{n+1}$ is defined in the 
predictive equation \eqref{NSTS.eq.pred.equation}
where $D_{n+1}  (\cdot)$ is   estimated by either $\hat D_{n+1}  (\cdot)$ 
\tr{
,   $ \hat D_{n+1}^{LLH}  (\cdot)$ 
or $ \hat D_{n+1}^{LLM}  (\cdot)$
}
\tr {all} with $T=t-1$. 

\item In practice, the integral \eqref{NSTS.eq.LMFpp} can be approximated by 
Monte Carlo, i.e.,   \\
\tr {$$\int g_{n+1}(x ) d\Phi(x) \simeq  \frac{1}{M} \sum_{i=1}^M g_{n+1}(x_i ) $$}
where $x_1,\ldots,x_M$ are generated as i.i.d.~$N(0,1)$, 
and $M$ is some large integer. 
\item Using the above 
Monte Carlo framework, the LMF $L_1$--optimal point predictor  of 
 $Y_{n+1} $ can be approximated by the median of the set
$\{  g_{n+1}(x_i )$ for $i=1,\ldots, M\}$.

\item 
Prediction intervals for $Y_{n+1} $ with prespecified coverage
probability can be constructed via the LMF Boootstrap of  
Algorithm \ref{MF3short.Algorithm2} 
based on either the $L_2$-- or $L_1$--optimal point predictor.
\end{enumerate}
\end{Algorithm} 
\vskip .13in
 \begin{Remark} \rm 
Interestingly, there is a closed-form solution for the LMF $L_1$--optimal point predictor  of  $Y_{n+1} $ that can also be used in 
  Step 5 of Algorithm \ref{NSTS.Algo.BasicMF}. To elaborate, first note that 
under the assumed weak dependence, e.g. strong mixing, of the series $\{Y_t\}$ 
(and therefore also of $\{Z_t\}$), we have the following approximations
(for large $n$), namely: 
$$Median \left(Z_{n+1} | {\cal F}_1^n(Z) \right)\simeq
Median \left(Z_{n+1} | {\cal F}_{-\infty} ^n(Z) \right)
$$ $$=
Median \left(Z_{n+1} | {\cal F}_{-\infty}^n(Y) \right)\simeq
Median \left(Z_{n+1} | {\cal F}_1^n(Y) \right).
$$
Now  eq.~\eqref{NSTS.eq.invtrans2} and \eqref{NSTS.eq.invtrans3}  imply 
that $Y_{n+1} =   D_{n+1}^{-1} \left(\Phi(Z_{n+1}) \right). $
Since $D_{n+1}(\cdot)$ and $\Phi (\cdot)$ are strictly increasing functions,
it follows that the Model-free $L_1$--optimal predictor of  $Y_{n+1}$ equals 
$$ Median \left(Y_{n+1} | {\cal F}_1^n(Y) \right) = 
  D_{n+1}^{-1} \left(\Phi \left( 
Median \left(Z_{n+1} | {\cal F}_1^n(Y) \right)
 \right)   \right)
$$
\begin{equation}
\label{NSTS.eq.invtrans5}
\simeq  D_{n+1}^{-1} \left(\Phi \left( 
Median \left(Z_{n+1} | {\cal F}_1^n(Z) \right)
 \right)   \right)
= D_{n+1}^{-1} \left(\Phi \left( 
E \left(Z_{n+1} | {\cal F}_1^n(Z) \right)
 \right)   \right)  ,
\end{equation}
the latter being due to the symmetry of the   normal distribution of $Z_{n+1} $ given $ {\cal F}_1^n(Z)$. 
But, as in eq.~\eqref{NSTS.eq.opt}, we have 
 $E \left(Z_{n+1} | {\cal F}_1^n(Z) \right) =
\phi_{1}(n) Z_n + \phi_{2}(n) Z_{n-1} + \ldots + \phi_{n}(n) Z_1$
where $   (\phi_{1}(n) , \cdots, \phi_{n}(n))'= \Gamma_n^{-1}  \gamma(n)$. Plugging-in  either $\bar D_{n+1}  (\cdot)$ 
\tr
{,  $ \bar D_{n+1}^{LLH}  (\cdot)$ 
or $ \bar D_{n+1}^{LLM}  (\cdot)$ 
}
in place of 
$D_{n+1}  (\cdot)$ in eq.~\eqref{NSTS.eq.invtrans5},
and also employing 
consistent estimates of $\Gamma_n$ and $ \gamma(n)$
 completes the calculation. As discussed in 
Section   \ref{NSTS.seq.whiteningtransformation}, $\Gamma_n$ can be estimated
by either $\hat \Gamma_n^{AR}$ or by the positive definite
banded estimator $\hat \Gamma_n^\star$ with a corresponding 
estimator for $ \gamma(n)$; 
\tr {see \cite{mcmurry2015high} for details.}
\end{Remark}

\begin{Remark} [Robustness of LMF approach] \rm
The   LMF approach focuses completely on the 
predictive equation~\eqref{NSTS.eq.pred.equation} for which an
estimate of (the inverse of) $D_{n+1} (\cdot)$ must be provided; 
interestingly, 
estimating $D_t(y)$ for $t\neq n+1$ is nowhere used 
in Algorithm \ref{NSTS.Algo.LMF}. In the usual case where 
the kernel $K(\cdot)$ is chosen to have compact support, 
estimating $D_{n+1} (\cdot)$ is only based on the last $b$ data values
$Y_{n-b+1},\ldots, Y_n$. Hence, in order for the LMF 
Algorithm \ref{NSTS.Algo.LMF} to be valid, the sole requirement
is that the subseries $Y_{n-b+1},\ldots, Y_n,Y_{n+1}$ is 
approximately stationary.  In other words, the first (and biggest) 
part of the data, namely $Y_1,\ldots, Y_{n-b}$, can suffer from 
arbitrary nonstationarities, change points, outliers, 
etc.~{\it without  the LMF predictive inference for $Y_{n+1}$ being 
  affected; }
this robustness of the LMF approach is highly advantageous.

\end{Remark}

\subsection{Discrete-valued time series}
\label{NSTS.discrete data}

Untill now, it has been assumed   that $D_{t }(y)$ is (absolutely) continuous in $y$ for all $t$; in this subsection, we briefly discuss   a departure from this assumption.

Throughout subsection \ref{NSTS.discrete data} we will assume that
the locally stationary time series $\{Y_t\}$ takes values in a countable set 
$S\subset {\bf R}$;
as an example, consider the case of a finite state Markov chain whose
first marginal changes smooth (and smoothly) with time. It is apparent
that $D_t(y)$ is a step function; hence, step function estimators
such as $\hat D_{t}  (y)$ 
\tr
{,   $ \hat D_{t}^{LLH}  (y)$ 
or $ \hat D_{t}^{LLM}  (y)$
}
are preferable to their smoothed counterparts
$\bar D_{t}  (y)$
\tr{
,  $ \bar D_{t}^{LLH}  (y)$
or $ \bar D_{t}^{LLM}  (y)$
}
since the latter assign positive probabilities to values $y \not \in S$. 

Fortunately, the LMF methodology of Algorithm \ref{NSTS.Algo.LMF}
can be employed based on just the step function estimators
  $\hat D_{t}  (y)$ 
\tr{,   $ \hat D_{t}^{LLH}  (y)$
or $ \hat D_{t}^{LLM}  (y)$.
}
Note that with discrete data, predicting $Y_{n+1}$ 
by a conditional mean or median makes little sense since the
latter will likely not be in the  set 
$S $; it is more appropriate to adopt a 0-1 loss function
and predict $Y_{n+1}$ 
by the {\it mode} of the  conditional distribution. A  
  prediction interval is not appropriate either unless 
the set $S$ is of lattice form---and even then, problems
ensue regarding non-attainable $\alpha$--levels.  It is thus 
more informative to present an estimate of the conditional distribution 
instead of summarizing the latter into a prediction interval.

\tr {A version of the LMF algorithm for discrete valued data is given below; (for details \tb{see} \cite{politis2015model}.}

\vskip .17in
\begin{Algorithm} {\sc 
  LMF bootstrap for   predictive distribution of discrete-valued 
  $Y_{n+1}$}
\label{NSTS.Algorithm3}
\begin{enumerate}
\item Based on the data $\underline{Y}_n$, estimate the
inverse transformation  $H_n^{-1}$
by    $\hat H_n^{-1}$ (say).
   In addition, estimate $  g_{n+1}$  by $\hat g_{n+1}$.

\item
\begin{enumerate}

\item Generate  bootstrap pseudo-data $\varepsilon_1^{\ast }, ..., \varepsilon_n^{\ast}$  
as i.i.d.~from   $F=\Phi$.

\item Use the inverse transformation  $\hat H_n^{-1}$ to
create pseudo-data in the $Y$ domain, i.e.,
let $\underline{Y}_n^\ast=(Y_1^\ast ,...,Y_n^\ast )'=
\hat H_n^{-1} (\varepsilon_1^{\ast}, ..., \varepsilon_n^{\ast})$.

\item  
Based on the bootstrap pseudo-data $\underline{Y}_n^\ast$, re-estimate the
transformation  $H_n$ and its inverse $H_n^{-1}$
by $\hat H_n^*$ and  $\hat H_n^{-1*}$ respectively.
   In addition, re-estimate $  g_{n+1}$  by $\hat g_{n+1}^*$.

\item Calculate a  bootstrap pseudo-value 
$Y_{n+1}^{**}$
as the point
$  \hat g_{n+1}^* ( \underline{Y}_n  ,   \varepsilon   )
$ where $\varepsilon $ is
generated  from $F=\Phi$.

\end{enumerate}

\item Steps (a)---(d) in the above should be repeated 
$B$ times (for some large $B$), 
  and the $B$ bootstrap  
replicates of the  pseudo-values
$ Y_{n+1}^{**} $ are collected in the form of an empirical distribution 
which is our Model-free estimate of the 
predictive distribution of $ Y_{n+1} $; the {\em mode} of this
distribution is the LMF  optimal predictor of $ Y_{n+1} $ under 0-1 loss. 
 
\end{enumerate}
\end{Algorithm}
\vskip .17in
\subsection{Special case: strictly stationary data}
\label{NSTS.sec.strictly stationary data}

It is interesting to consider what happens if/when the data $Y_1,\ldots,Y_n$
are a stretch of a strictly stationary time series $\{Y_t\}$. 
Of course, a time series that is strictly stationary
is a {\it a fortiori} locally stationary; so all the aforementioned
procedures should work {\it verbatim}. Nevertheless, one could take
advantage of the stationarity to obtain better estimators;  
effectively,  one can take the bandwidth $b$ to be comparable to $n$,
i.e., employ global---as opposed to local---estimators.

To elaborate, in the stationary case the distribution $D_{t }(y)$ 
does not depend on $t$ at all. Hence, for the purposes of the 
LMF Algorithm  \ref{NSTS.Algo.LMF}---as well as the 
discrete data Algorithm~\ref{NSTS.Algorithm3}---we can estimate $D_{t }(y)$  by
the regular (non-local) empirical distribution 
$$\hat D (y)= n^{-1}\sum_{t=1}^n {\bf 1}\{ Y_t\leq y\}.$$
Furthermore, for the purposes of Algorithm \ref{NSTS.Algo.BasicMF}  
 we can estimate the (assumed smooth) $D_{t }(y)$   by
the smoothed empirical distribution 
$$\bar D (y)= n^{-1}\sum_{t=1}^n \Lambda(\frac {y-Y_{t}} {{h}_0}) $$
where ${h}_0$ is a positive bandwidth parameter satisfying ${h}_0\to 0$
as $n\to \infty$. As mentioned in 
\tr {Section \ref{NSTS.sec.MF.cv}},
the optimal rate is 
$h_0\sim n^{-2/5}$ when the estimand $D_{t }(y)$ is sufficiently
smooth in $y$.

\subsection{Local stationarity in a higher-dimensional marginal}
\label{NSTS.sec.higher-dimensional marginals}
The success of the  theoretical transformation of Section \ref{NSTS.sec.CTT}
in transforming the data vector
$ \underline{Y}_{n}$ to the vector of i.i.d.~components
   $\underline{\epsilon_{n}}$ hinges on two conditions:
(a) the nonstationarity of $\{Y_t\}$ is only due to nonstationarity 
in its first marginal $D_t(\cdot)$, and (b) the instantaneous 
transformation to Gaussianity also manages to create a Gaussian 
random vector, i.e., all its finite-dimensional marginals are Gaussian.
Both of these conditions can be empirically checked. For example,
condition (a)
can be checked by looking at some features of interest of 
the $m$th (say) marginal, e.g., looking at the autocorrelation
 Corr$ (Y_{t}, Y_{t+m})$ estimated over different subsamples of the
data, and checking whether it depends on $t$. Condition (b) can be 
checked by performing a normality test, e.g., Shapiro-Wilk test, 
or other diagnostics, e.g., quantile plot, on selected 
linear combinations of $m$ consecutive components of the random vector. 

Interestingly, if either   condition (a) or (b) seem to fail, there is a
single solution to address the problem, namely blocking the time series. 
 To elaborate,  one would then create blocks of data by 
defining $B_t=(Y_t,\ldots,Y_{t+m-1})' $ for $t=1,\ldots, q$ with $q=n-m+1$. 
Now focus on the multivariate time series dataset $\{B_1,\ldots, B_q\}$, and
let $D_t^{(m)}(\cdot) $ denote the distribution function of vector $B_t$ which
will be assumed to vary smoothly (and slowly) with $t$
as in Remark \ref{NSTS.re.smooth2}.

Using the 
\tr {\cite{rosenblatt1952remarks}}
transformation, we can now map $B_t$ to a random vector $V_t$
that has components\footnote{Recall that the 
\tr {\cite{rosenblatt1952remarks}}
transformation   maps an arbitrary random
vector $\underline{Y}_m=(Y_1,\ldots, Y_m)^\prime $ having  
absolutely continuous joint distribution onto 
a random vector  $\underline{V}_m=(V_1,\ldots, V_m)^\prime $ whose
entries are i.i.d.~Uniform(0,1); this is done via the probability integral transform based on  conditional distributions. To elaborate, for $k>1$ define 
the conditional distributions
$D_k(y_k|y_{k-1},\ldots, y_1)=
P \{ Y_k\leq y_k|Y_{k-1}=y_{k-1},\ldots, Y_1= y_1\}$, and let 
$D_1(y_1)= P \{ Y_1\leq y_1\}$.
Then, the 
\tr{\cite{rosenblatt1952remarks}}
transformation  amounts to letting
$V_1=D_1(Y_1), V_2=D_2(Y_2|Y_1), V_3=D_3(Y_3|Y_2,Y_1), \ldots , $ and 
$ V_m=D_m(Y_m|Y_{m-1}, \ldots, Y_2,Y_1)$.  }
  i.i.d.~Uniform (0,1),
and then do the Gaussian transformation and whitening as required by the Model-Free Principle.
 Thus, when the  
time series $\{Y_t\}$   is locally stationary in its $m$th marginal,
the   algorithm  to transform the dataset $\underline{Y}_n=(Y_1,\ldots, Y_n)^\prime $ to an i.i.d.~dataset goes as follows.

\vskip .17in

\begin{enumerate}
\item From the dataset $\underline{Y}_n=(Y_1,\ldots, Y_n)^\prime $, create  blocks/vectors  $B_t=(Y_t,\ldots,Y_{t+m-1})' $ for $t=1,\ldots, q$ with $q=n-m+1$.
\item Use the Rosenblatt  
transformation to map  the multivariate   dataset $\{B_1,\ldots, B_q\}$
to the dataset $\{V_1,\ldots, V_q\}$; here $V_t=
(V_t^{(1)}, \ldots , V_t^{(m)})'$ is a 
 random vector having   components   that
are   i.i.d.~Uniform (0,1). 
\item Let $Z_t^{(j)}=\Phi ^{-1}( V_t^{(j)})$ for
$j=1, \ldots, m$, and $t=1,\ldots, q $ where $\Phi$ is the cdf of a standard normal. Note that, for each $t$, 
  the variables $Z_t^{(1)}, \ldots , Z_t^{(m)}$ are i.i.d.~$N(0,1)$. 

\item 
Define the   vector time series $Z_t= ( Z_t^{(1)}, \ldots , Z_t^{(m)})'$
that is multivariate Gaussian. Estimate the (matrix) 
autocovariance sequence Cov$(Z_t,Z_{t+k})$ for
$k=0,1,\ldots$, and
use it to   `whiten'  the sequence $ Z_1,\ldots, Z_q $, i.e., to map it
(in a one-to-one way) to the i.i.d.~sequence $ \zeta_1,\ldots, \zeta_q $;
here,
  $\zeta_t\in {\bf R}^m$ is a random vector having   components   that
are   i.i.d.~$N(0,1)$. 
\end{enumerate}
  
\vskip .1in
\noindent
In   Step 2   above, the $m$th dimensional Rosenblatt  
transformation  can be estimated in practice using 
a local average or local linear estimator, i.e., 
a multivariate analog of $\bar D_t(\cdot)$ 
\tr 
{, 
$\bar D_t^{LLH}(\cdot)$
or $\bar D_t^{LLM}(\cdot)$
.}
 Regarding Step 4, standard methods  exist to estimate the (matrix) 
autocovariance of $Z_t$ with $Z_{t+k}$; see e.g. 
\tr{\cite{jentsch2015covariance}.}
Finally, note that the map $H_n: \underline{Y}_n \mapsto 
(\zeta_1,\ldots, \zeta_q)'$ is invertible
since all four steps given above are one-to-one. Hence, Model-free
prediction can take place based on a 
multivariate version of the Model-free Prediction Principle
of 
\tr {\cite{Politis2013}};
the details are straightforward.

\section{\tr {Diagnostics for Model-Free Inference}}
\label{NSTS.Diagnostics}

\tr {The steps outlined in Section \ref{NSTS.sec.CTT} for Model-Free inference involve generating samples from both uniform
$U[0,1]$ and standard normal distributions. Careful analysis is necessary to ensure that the samples generated are from the correct 
distributions failing which the Model-Free point and interval predictors will be inaccurate. The following discussion serves
as an aid to the practitioner to ensure realization of optimal performance for both point prediction and prediction interval
generation using the Model-Free methodology.}

\subsection{\tr{QQ-plots after uniformization}}
\label{MF.qq_plot}

\tr{The success of the uniformization step outlined in Section \ref{NSTS.sec.CTT} can be visually verified using QQ-plots
of the obtained uniform samples versus samples obtained from an ideal uniform distribution which is available in
standard statistical software such as R. Any deviations in these curves from linearity should be closely investigated for possible
issues wrt choice of bandwidth during cross-validation as it can impact both point prediction and prediction interval
generation.}

\subsection{\tr{Shapiro-Wilk test for joint normality}}
\label{MF.Shapiro}

\tr{The random vector $\underline{Z}_n=(Z_1,\ldots, Z_n)'$ \tb{from Section \ref{NSTS.seq.whiteningtransformation}} should be tested for normality in order to ensure that
the \tb{described} whitening transformation successfully produces i.i.d. normal
samples. \tb{Marginal} normality of the data $Z$ can be verified \tb{by gauging} linearity of QQ-plots versus the standard normal distribution. \tb{Furthermore} the Cramer-Wold theorem  states that any linear combination of jointly normal variables is univariate normal. This can be  used to empirically verify whether the joint normality requirement is violated by taking any linear combination
i.e.  for example a pair or triplet of variables from the set $\underline{Z}_n=(Z_1,\ldots, Z_n)'$ and verify their normality using the
Shapiro-Wilk test.  An example of this is provided in Figure \ref{shap_wilk}  where for a given $\lambda$ we form the linear
combination \tb{$(1-\lambda)Z_{i} + \lambda Z_{i+1}$}  over all obtained values $\underline{Z}_n=(Z_1,\ldots, Z_n)'$ and calculate the mean value
of the Shapiro-Wilk test statistic.  This is done over a range of $\lambda$ values. As can be seen from the plot sufficiently
high values of the test statistic are obtained which indicates that from this particular test we cannot conclude that joint
normality has been violated. Further tests can be done by forming linear combinations over pairs of non-successive
values of $Z$.
 }

\subsection{\tr{Kolomogorov-Smirnov test for i.i.d. standard normal samples}}
\label{MF.KS}

\tr{Provided that the inputs are jointly normal the whitening transformation described in \tb{Section} \ref{NSTS.seq.whiteningtransformation}
produces i.i.d. standard normal variables. The covariance matrix used in this step can be derived either by fitting a causal AR(p)
model to $\underline{Z}_n=(Z_1,\ldots, Z_n)'$ or using the flat-top kernel banded, tapered estimator outlined in
\cite{mcmurry2010banded}. To verify that the data generated after whitening are standard normal a Kolmogorov-Smirnov test can be 
used with the reference distribution as $N[0,1]$.}

\subsection{\tr{Independence test of standard normal samples}}
\label{MF.ind}

\tr{The success of the Model-Free procedure involves the ability to produce i.i.d. data after a series of invertible transformations.
In the case of \tb{Locally Stationary Time Series} independence of the data produced at the final step after applying the whitening transformation can be verified
visually using an autocorrelation function (ACF) plot as the data are \tb{approximately} standard normal. An example of this is given in Figure \ref{ACF_lag_100}
where it can be noticed from the ACF plot that the \tb{Model-Free} transformations were successful in producing decorrelated and therefore i.i.d. (normal) data.}

\graphicspath{{/Users/rumpagiri/Documents/NONPARAMETRIC/model_free/papers/LSTS}}
\DeclareGraphicsExtensions{.png}

{\begin{figure}[!t]
  \centering
  \includegraphics[width=3.5in, height=2.5in]{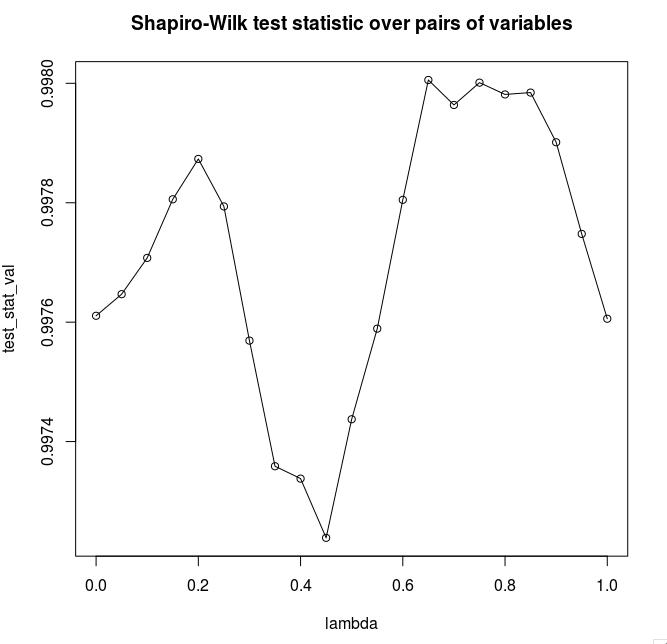}
  \caption{Values of Shapiro-Wilk test statistic for joint normality test. \tb{Note that corresponding p-values range from 0.09 to 0.29.}}
  \label{shap_wilk}
\end{figure}}

\graphicspath{{/Users/rumpagiri/Documents/NONPARAMETRIC/model_free/papers/LSTS}}
\DeclareGraphicsExtensions{.png}

{\begin{figure}[!t]
  \centering
  \includegraphics[width=3.5in, height=2.5in]{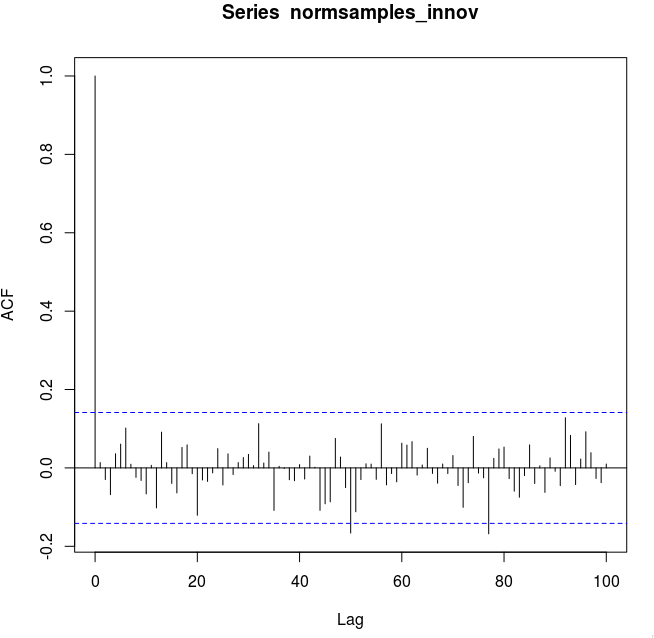}
  \caption{Autocorrelation plot showing decorrelation/independence of data after whitening transformation}
  \label{ACF_lag_100}
\end{figure}}

\section{Model-Free vs. Model-Based Inference: 
 \tr {empirical comparisons}}
\label{NSTS.Numerical}

\tr {
The performance of the Model-Free and Model-Based \tb{predictors} described above are empirically compared using both simulated and real-life datasets based on point prediction and also calculation of prediction intervals.  The Model-Based local constant and local linear methods are denoted as \tb{MB-LC} and \tb{MB-LL} respectively. \tb{Model-Based predictors MB-LC and MB-LL are described in Section \ref{NSTS.Model-based inference}}. \tbl{The Model-Free methods using local constant, local linear (Hansen) and local linear (Monotone) using the flat-top tapered covariance estimator are denoted as MF-LC, MF-LLH, MF-LLM.} \tbl{Model-Free methods using local constant, local linear (Hansen) and local linear (Monotone) \tbl{using the covariance estimator obtained from fitting a causal AR(p) model} are denoted as MF-LC-ARMA, MF-LLH-ARMA, MF-LLM-ARMA. Model-Free predictors are described in Section \ref{NSTS.Model-free inference}. The covariance estimators using the flat-top tapered kernel and fitting an AR(p) model are discussed in Section \ref{NSTS.seq.whiteningtransformation}.} Results are also shown for the LMF counterparts of these methods which are denoted as LMF-LC, LMF-LLH, LMF-LLM \tbl{and LMF-LC-ARMA, LMF-LLH-ARMA, LMF-LLM-ARMA respectively}. Results for all methods are given for both fitted (F) and predictive (P) residuals.
Following metrics are used to compare the estimators:
}
\begin{enumerate}
\item
\tr {
Point prediction performance as indicated by Bias and Mean Squared Error (MSE) on simulated and real-life datasets
using all Model-Based and Model-Free methods listed above.
}
\item
\tr {
Bootstrap performance as indicated by coverage probability (CVR), mean length of prediction intervals and standard
deviation (sd) of length of prediction intervals. All prediction interval metrics given in the following tables have been
generated \tb{based on} a nominal coverage of $90\%$.
}
\end{enumerate}

\subsection{Simulation: Additive model with stationary  AR(5) errors  }

\tr{Data $Y_i$ for $t =1, \ldots, 1000$ were simulated as per model \eqref{NSTS.eq.model homo}
with trend as in eq.~\eqref{NSTS.eq.qs},
i.e., $\mu (t)= \mu_{_{[0,1]}} (a_t)$ 
with $a_t= (t-1)/n$ and $\mu_{_{[0,1]}} (x)= \sin (2\pi x)$. 
The series $W_t$ is constructed  via 
an AR(5)   model  driven by errors $ V_t$ that 
are i.i.d.~$N(0, \tau^2);$ with $\tau=0.14$.  The AR(5)
 coefficients
 are set to 0.5, 0.1, 0.1, 0.1, 0.1.
Sample size $n$ is set to $1000$. Point prediction and prediction intervals
 are measured for boundary point $n = 1000$. Bandwidths for estimating the trend are calculated
 using the cross-validation techniques for Model-Based and Model-Free cases described in Sections
\ref{NSTS.sec.trend} and \ref{NSTS.sec.MF.cv} respectively.}

\tr {Results for point prediction including bias and mean square error (MSE) over all MB and MF methods are shown in Table \ref{pp_AR_5} below.
A total of 500 realizations of the dataset were used for measuring point prediction performance.}

\tr {Results for prediction intervals including CVR, length and standard deviation of
the predicted intervals over all MB and MF methods are shown in Table \ref{bt_AR_5} below. A total of 250 realizations were used
for measuring prediction interval performance. The number of bootstrap replications $B$ was set to 250.}


\tbl{
From point-prediction results on this dataset it can be seen that one of the best predictors is MB-LL; this is expected since the LL regression estimator
is great for extrapolation, and  the  innovations are generated using an AR model which is  directly employed in the MB-LL estimator. 
Nevertheless, predictors  MF-LLM and MF-LLM-ARMA appear equally as
good which is re-assuring and surprising at the same time;
it appears that---as with the case of regression with independent errors \cite{das2017nonparametric}---the monotonicity correction in the LLM distribution estimator has minimal effect on the center of the distribution that is used for
point prediction. The MF-ARMA and LMF-ARMA outperform their respective MF and LMF counterparts for point prediction; this is consistent with that fact that the data is generated by an AR process and therefore the covariance estimator using AR($p$) estimation outperforms its flat-top tapered counterpart. However the  MF-LLM,  LMF-LLM, MF-LLM-ARMA and LMF-LLM-ARMA estimators give the best prediction intervals when both coverage probabilities and mean interval lengths are considered. This is a somewhat surprising result given the fact that the data was generated using an AR(5) model, and   one would expect that the model-based estimator MB-LL would perform comparably with its MF counterparts, i.e., MF-LLM and MF-LLM-ARMA, in terms  of prediction intervals.
}

\tbl{
Among the MF estimators it is the MF-LLM, LMF-LLM, MF-LLM-ARMA and LMF-LLM-ARMA methods that perform better than their LC and LLH counterparts both for the flat-top tapered and AR(p) based covariance estimators.} This improvement can be attributed to using negative weights for estimation at the boundary with
the Monotone Local Linear Distribution estimator i.e. the LLM methods.

\tr{
\tb{As before} prediction interval coverage is \tb{enhanced} using predictive as compared to fitted residuals which is consistent with the results
of interval coverage using both types of residuals as discussed for the regression case in \cite{Politis2013}.
}

\bigskip

\begin{table}
\centering
\caption{Point Prediction performance for AR(5) dataset}
\label{pp_AR_5}
\begin{tabular}{|w|w|w|w|}
\hline
Prediction \ Method  & Residual Type & Bias & MSE\\
\hline
MB-LC & P & {-2.899e-02} & {2.878e-02}\\
\hline
& F & {-3.310e-02} & {2.923e-02}\\
\hline
MB-LL & P & {-3.031e-03} & {2.848e-02}\\
\hline
& F & {-7.315e-03} & {2.841e-02}\\
\hline
MF-LC & P & {-3.910e-02} & {2.955e-02}\\
\hline
& F & {4.327e-02} & {2.949e-02}\\
\hline
MF-LLH & P & {-3.591e-02} & {2.996e-02}\\
\hline
& F & {-4.177e-02} & {3.000e-02}\\
\hline
MF-LLM & P & {-2.716e-02} & {2.832e-02}\\
\hline
& F & {-3.599e-02} & {2.909e-02}\\
\hline
LMF-LC & P &  {-3.915e-02} & {2.961e-02}\\
\hline
& F & {-4.349e-02} & {2.953e-02}\\
\hline
LMF-LLH & P & {-3.691e-02} & {2.996e-02}\\
\hline
& F & {-4.224e-02} & {3.010e-02}\\
\hline
LMF-LLM & P & {-2.753e-02} & {2.855e-02}\\
\hline
& F & {-3.614e-02} & {2.915e-02}\\
\hline
MF-LC-ARMA & P & {-3.418e-02} & {2.929e-02}\\
\hline
& F & {-3.932e-02} & {2.920e-02}\\
\hline
MF-LLH-ARMA & P & {-3.067e-02} & {2.941e-02}\\
\hline
& F & {-3.766e-02} & {2.917e-02}\\
\hline
MF-LLM-ARMA & P & {-2.226e-02} & {2.829e-02}\\
\hline
& F & {-3.219e-02} & {2.876e-02}\\
\hline
LMF-LC-ARMA & P & {-3.452e-02} & {2.957e-02}\\
\hline
& F & {-3.968e-02} & {2.942e-02}\\
\hline
LMF-LLH-ARMA & P & {-3.141e-02} & {2.942e-02}\\
\hline
& F & {-3.776e-02} & {2.927e-02}\\
\hline
LMF-LLM-ARMA & P & {-2.229e-02} & {2.824e-02}\\
\hline
&F & {-3.300e-02} & {2.893e-02}\\
\hline
\end{tabular}
\end{table}

\begin{table}
\centering
\caption{Interval estimation performance using bootstrap for AR(5) dataset}
\label{bt_AR_5}
\begin{tabular}{| w | w | w | w | w |}
  \hline
  Prediction Method  & Residual Type & CVR & Mean Length & SD Length\\
  \hline
  MB-LC & P & {0.88} & {7.001e-01} & {1.781e-01} \\
  \hline
   & F & {0.83} & {5.598e-01} & {2.013e-01}\\
  \hline
  MB-LL & P & {0.92} & {7.802e-01} & {1.718e-01}\\
  \hline
   & F & {0.88} & {7.039e-01} & {1.725e-01}\\
\hline
   MF-LC & P & {0.85} & {7.443e-01} & {1.500e-01} \\
  \hline
    & F & {0.83} & {6.362e-01} & {1.709e-01} \\
  \hline
  MF-LLH & P & {0.88} & {7.489e-01} & {1.422e-01}\\
  \hline
   & F & {0.84} & {6.495e-01} & {1.234e-01}\\
  \hline
   MF-LLM & P & {0.89} & {7.343e-01} & {1.386e-01}\\
  \hline
   & F & {0.88} & {6.422e-01} & {1.229e-01}\\
  \hline
   LMF-LC & P & {0.86} & {7.424e-01} & {1.515e-01}\\
  \hline
   & F & {0.83} & {6.373e-01} & {1.492e-01}\\
  \hline
  LMF-LLH & P & {0.88} & {7.582e-01} & {1.386e-01}\\
  \hline
   & F & {0.85} & {6.534e-01} & {1.275e-01}\\
  \hline
   LMF-LLM & P & {0.89} & {7.423e-01} & {1.401e-01}\\
  \hline
   & F & {0.88} & {6.460e-01} & {1.278e-01}\\
  \hline
   MF-LC-ARMA & P & {0.85} & {7.452e-01} & {1.485e-01}\\
  \hline
    & F & {0.80} & {6.317e-01} & {1.421e-01}\\
  \hline
   MF-LLH-ARMA & P & {0.85} & {7.474e-01} & {1.416e-01}\\
  \hline
    & F & {0.84} & {6.569e-01} & {1.286e-01}\\
  \hline
    MF-LLM-ARMA & P & {0.88} & {7.362e-01} & {1.442e-01}\\
  \hline
    &F & {0.87} & {6.502e-01} & {1.264e-01}\\
  \hline
     LMF-LC-ARMA & P & {0.85} & {7.437e-01} & {1.485e-01}\\
  \hline
    & F & {0.82} & {6.382e-01} & {1.452e-01}\\
  \hline
  LMF-LLH-ARMA & P & {0.86} & {7.428e-01} & {1.389e-01}\\
  \hline
   & F & {0.85} & {6.564e-01} & {1.254e-01}\\
  \hline
   LMF-LLM-ARMA & P & {0.88} & {7.422e-01} & {1.423e-01}\\
  \hline
   & F & {0.87} & {6.519e-01} & {1.278e-01}\\
  \hline
  \end{tabular}
\end{table}
 

\subsection {\tr {Simulation: Additive model with nonlinearly generated errors}}

\tr{Data $Y_i$ for $t =1, \ldots, 1000$ were simulated from model \eqref{NSTS.eq.model homo}
with trend as in eq.~\eqref{NSTS.eq.qs},
i.e., $\mu (t)= \mu_{_{[0,1]}} (a_t)$ 
with $a_t= (t-1)/n$ and $\mu_{_{[0,1]}} (x)= 5*\sin (2\pi x)$. }
\tbl {The series  $W_t$ is now constructed via the nonlinear model given below}:


\begin{align}
\label{NSTS.eq.nlarma}
\tr {W_{t} = \begin{cases}
                 1 + \alpha{W_{t-1}}  + e_t & \tb{if} \ W_{t-1}  \leq r \\
                -1 + \beta{W_{t-1}} + \gamma{e_t} & \tb{if} \ W_{t-1}  > r
                 \end{cases}}
\end{align}
\noindent
\tbl {where the errors  $ e_t$    are assumed i.i.d.~$N(0, \tau^2)$. 
Eq.  (\ref{NSTS.eq.nlarma}) describes a TAR(1) model, i.e., Threshold
Autoregression of order 1; see \cite{tong2011threshold} and the references therein. For our implementation, we chose $\tau=0.4$, $\alpha=0.5$, $\beta=-0.6$, $r=0.6$, $\gamma=1$; the initial value of $W_t$ is set to 0, and $n=$  $1000$. 
A scatterplot  showing $W_{t}$ versus $W_{t-1}$ is shown in Figure \ref{nl_innov}. The process of eq. (\ref{NSTS.eq.nlarma}) is not zero-mean; however its mean is  removed during detrending either with   Model-Based or Model-Free methods.}
Point prediction and prediction intervals are measured for boundary point $n = 1000$. Bandwidths for estimating the trend are calculated
 using the cross-validation techniques for Model-Based and Model-Free cases described in Sections
\ref{NSTS.sec.trend} and \ref{NSTS.sec.MF.cv} respectively.

\tr {Results for point prediction including bias and mean square error (MSE) over all MB and MF methods are shown in Table \ref{pp_NL} below.
A total of 500 realizations of the dataset were used for measuring point prediction performance.}

\tr {Results for prediction intervals including CVR, length and standard deviation of
the predicted intervals over all MB and MF methods are shown in Table \ref{bt_NL} below. A total of 250 realizations were used
for measuring prediction interval performance. The number of bootstrap replications $B$ was set to 250.}

\tbl{
From point-prediction results on this dataset it can be seen that the MF-LLM-ARMA and LMF-LLM-ARMA estimators give the best performance. The MF-ARMA and LMF-ARMA outperform their respective MF and LMF counterparts for point prediction. This is consistent with that fact that the data is not generated by an MA process and therefore the covariance estimator using AR(p) estimation outperforms its flat-top tapered counterpart which assumes an MA model. The MF-LLM, LMF-LLM, MF-LLM-ARMA and LMF-LLM-ARMA estimators give the best prediction intervals when both coverage probabilities and mean interval lengths are considered. These results are somewhat expected since the innovations are generated using a nonlinear model and the MB methods use a linear predictor.}
Therefore MF-LLM and LMF-LLM estimators perform better than their model-based counterparts i.e. the MB-LL methods.
\tb {However it is striking to see a Model-Free method outperform the Model-Based ones when the additive model is true.}

\tr{
\tbl{It can also be seen that for most cases} prediction interval coverage is \tb{enhanced} using predictive as compared to fitted residuals which is consistent with the results
of interval coverage using both types of residuals as discussed for the regression case in \cite{Politis2013}.
}

\graphicspath{{/Users/rumpagiri/Documents/NONPARAMETRIC/model_free/papers/LSTS}}
\DeclareGraphicsExtensions{.png}

{\begin{figure}[!t]
  \centering
  \includegraphics[width=3.5in, height=2.5in]{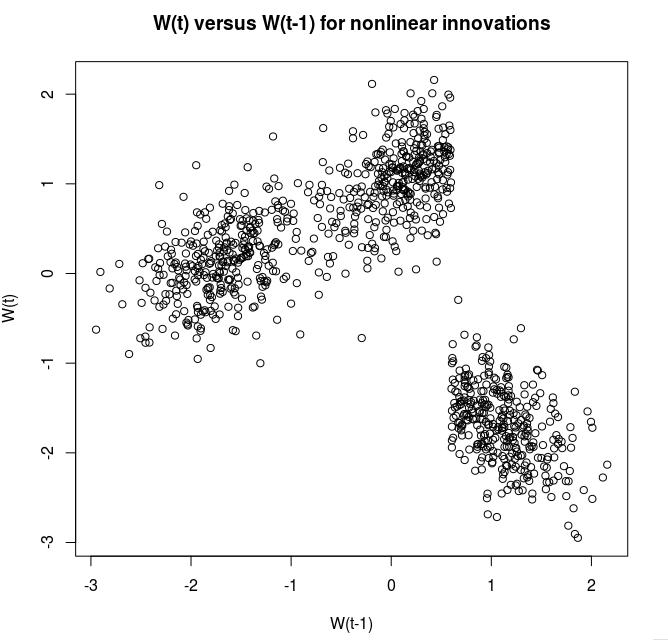}
  \caption{Nonlinear time series scatterplot of $W_{t}$ versus $W_{t-1}$.}
  \label{nl_innov}
\end{figure}}


\begin{table}
\centering
\caption{Point Prediction performance for nonlinear  dataset}
\label{pp_NL}
\begin{tabular}{|w|w|w|w|}
\hline
Prediction \ Method  & Residual Type & Bias & MSE\\
\hline
{MB-LC} & P &{-1.894e-01} & {8.420e-01}\\
\hline
& F &  {-1.897e-01} & {8.542e-01}\\
\hline
{MB-LL} & P & {-1.109e-01}& {8.003e-01}\\
\hline
& F & {-1.082e-01} & {8.048e-01}\\
\hline
MF-LC & P  & {-1.697e-01} & {8.616e-01}\\
\hline
& F & {-1.937e-01} & {8.407e-01}\\
\hline
MF-LLH & P & {-1.134e-01} & {8.345e-01}\\
\hline
& F & {-1.193e-01} & {8.137e-01} \\
\hline
{MF-LLM} & {P} & {-2.418e-02} & {8.208e-01}\\
\hline
& F & {-1.770e-02} & {7.886e-01}\\
\hline
LMF-LC & P &  {-1.631e-01} & {8.671e-01}\\
\hline
& F & {-1.858e-01} & {8.456e-01}\\
\hline
LMF-LLH & P & {-1.004e-01} & {8.338e-01}\\
\hline
& F & {-1.108e-01} & {8.420e-01}\\
\hline
LMF-LLM & P & {-1.339e-02} & {8.287e-01}\\
\hline
& F & {-8.603e-03} & {7.941e-01}\\
\hline
MF-LC-ARMA & P  & {-1.151e-01} & {8.233e-01}\\
\hline
& F & {-1.308e-01} & {8.003e-01}\\
\hline
MF-LLH-ARMA & P & {-1.346e-01} & {8.075e-01}\\
\hline
& F & {-1.370e-01} & {7.945e-01} \\
\hline
{MF-LLM-ARMA} & {P} & {-9.632e-03} & {7.861e-01}\\
\hline
& F & {-5.183e-03} & {7.849e-01}\\
\hline
LMF-LC-ARMA & P &  {-1.214e-01} & {8.290e-01}\\
\hline
& F & {-1.390e-01} & {8.140e-01}\\
\hline
LMF-LLH-ARMA & P & {-1.274e-01} & {8.225e-01}\\
\hline
& F & {-1.340e-01} & {8.008e-01}\\
\hline
LMF-LLM-ARMA & P & {-4.025e-03} & {7.945e-01}\\
\hline
& F & {2.181e-03} & {7.966e-01}\\
\hline
\end{tabular}
\end{table}

\begin{table}
\centering
\caption{Interval estimation performance using bootstrap for nonlinear  dataset}
\label{bt_NL}
\begin{tabular}{| w | w | w | w | w |}
  \hline
  Prediction Method  & Residual Type & CVR & Mean Length & SD Length\\
  \hline
   MB-LC & P  & {0.86} & {3.265} & {3.864e-01}\\
  \hline
     & F & {0.81} & {2.837} & {3.841e-01}\\
  \hline
   {MB-LL} & P & {0.85} & {3.123} &  {3.383e-01} \\
  \hline
     & F & {0.81} & {2.780} &  {3.466e-01} \\
\hline
     {MF-LC} & P & {0.88} & {3.999} &  {5.874e-01} \\
  \hline
      & F &  {0.90} & {2.954} & {4.272e-01} \\
  \hline
  {MF-LLH} & P & {0.88} & {4.051} &  {6.745e-01}\\
  \hline
    & F & {0.84} & {2.732} &  {4.605e-01} \\
  \hline
     {MF-LLM} & {P} & {0.89} & {3.891} & {6.956e-01} \\
  \hline
     & {F} & {0.86} & {2.657} & {4.726e-01} \\
  \hline
     {LMF-LC} & P & {0.87} & {3.987} & {6.052e-01} \\
  \hline
     & F & {0.88} & {2.942} & {4.133e-01} \\
  \hline
  {LMF-LLH} & P & {0.88} & {4.042} & {6.797e-01}\\
  \hline
     & F & {0.84} & {2.723} & {4.373e-01} \\
  \hline
   {LMF-LLM} & P & {0.88} & {3.946} &  {6.620e-01} \\
  \hline
     & F & {0.84} & {2.661} &  {4.558e-01} \\
  \hline
    {MF-LC-ARMA} & P & {0.86} & {3.850} &  {5.307e-01} \\
  \hline
      & F &  {0.89} & {2.896} & {4.343e-01} \\
  \hline
  {MF-LLH-ARMA} & P & {0.89} & {3.917} &  {6.602e-01}\\
  \hline
     & F & {0.88} & {2.694} &  {4.719e-01} \\
  \hline
     {MF-LLM-ARMA} & {P} & {0.86} & {3.794} & {6.319e-01} \\
  \hline
     & {F} & {0.85} & {2.614} & {4.766e-01} \\
  \hline
    {LMF-LC-ARMA} & P & {0.88} & {3.981e} & {5.723e-01} \\
  \hline
   & F & {0.89} & {2.966} & {4.423e-01} \\
  \hline
   {LMF-LLH-ARMA} & P & {0.90} & {4.022} & {6.889e-01}\\
  \hline
     & F & {0.86} & {2.764} & {4.451e-01} \\
  \hline
     {LMF-LLM-ARMA} & P & {0.88} & {3.948} &  {6.556e-01} \\
  \hline
   & F & {0.86} & {2.659} &  {4.844e-01} \\
  \hline
    \end{tabular}
\end{table}

\section {\tr {Real-life example: Speleothem data}}
\label{NSTS.Speleothem}

\tor{
The Speleothem dataset first discussed in \cite{fleitmann2003holocene} and} \tc{further analyzed} \tor{in \cite{mudelsee2014climate} is an interesting real-life example to compare metrics of point prediction and prediction intervals for all MB and MF estimators described before. This dataset which is shown in Figure \ref{Mudelsee_full} contains oxygen isotope record (the ratio of $^{18}O$ to $^{16}O$) from stalagmite Q5 from southern Oman over the past 10,300 years. The oxygen isotope ratio obtained from the speleothem climate archive serves as a proxy variable for the actual climate variable {\bf monsoon rainfall}.  The full dataset has $Y_i$ for $t =1, \ldots, 1345$ points which are in general \tc{obtained} with unequal spacing. The following points should be noted in the context of our analysis of the speleothem proxy dataset:}
\begin{enumerate}

\item 
\tor{
One important application of proxy data obtained from climate archives is prediction of the unobserved climate variable values. This prediction is based on known values of proxy and climate variables which in this case are the oxygen isotope ratio and monsoon rainfall respectively. Proxy data  are also useful for construction of confidence intervals for parameter estimates of the proxy variable model. In our case we use a part of the proxy variable dataset which contains a linear trend for estimating the performance of Model-Based and Model-Free predictors for the proxy variable delta-O-18.}

\item

Proxy data obtained from climate archives may be obtained over either even or uneven time spacing. In case of the speleothem dataset under consideration as shown in Figure \ref{Mudelsee_age} the spacing variations are \tc{small in general and definitely} negligible over the part of the dataset (last 62 points) where we perform prediction; \tnb{see Figure \ref{Mudelsee_age} that depicts the age versus sample number}. Hence we will assume even time spacing in our analysis. No interpolation is applied i.e. the number of time-points assumed with even spacing is the same as the number of time points which are present with slightly uneven spacing in the original dataset. It is to be noted that several other techniques such as Singular Spectrum Analysis, Principal Component Analysis and Wavelet Analysis also assume even spacing for time-series analysis. Extension of our methods to incorporate uneven time spacing will be the focus of future work.

\end{enumerate}

\graphicspath{{/Users/rumpagiri/Documents/NONPARAMETRIC/model_free/papers/LSTS}}
\DeclareGraphicsExtensions{.png}

{\begin{figure}[!t]
  \centering
  \includegraphics[width=3.5in, height=2.5in]{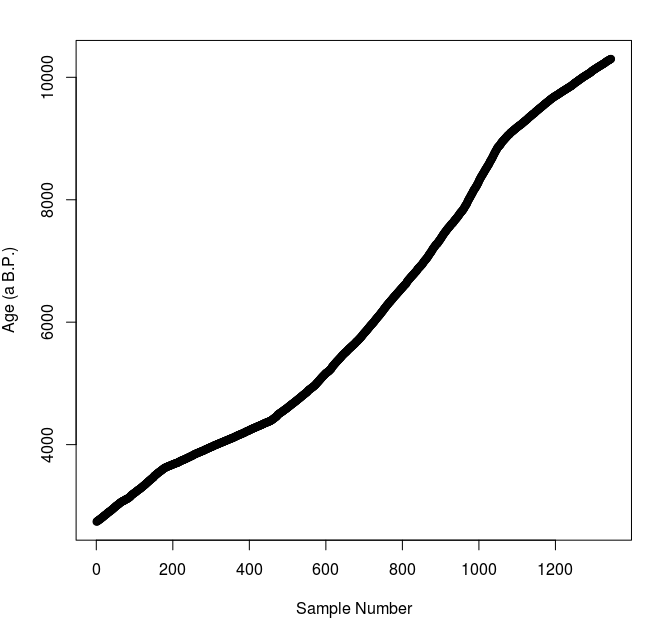}
  \caption{Age (a B.P.) of delta-O-18 versus sample number}
  \label{Mudelsee_age}
\end{figure}
}

\tnr{
We consider the dataset over the last 270 points as shown in Figure \ref{Mudelsee_short}. This dataset is divided into 2 parts: the first part is used to determine the bandwidths for the MB and MF  estimators using methods outlined in Sections \ref{NSTS.sec.trend} and \ref{NSTS.sec.MF.cv} respectively; the last  62 points are used to calculate point prediction and prediction intervals. It can be noticed from Figures \ref{Mudelsee_full} and \ref{Mudelsee_short} that this last part of the data appears to have a linear trend. A moving window method is adopted for cross-validation i.e. for point $Y_t$ (whose metrics for point prediction and prediction intervals are calculated) we use points [$Y_{t-w}$, $Y_{t-1}$] for cross-validation. Here the value of $w$ is set to 189. Note also that since this dataset contains a smaller number of points, cross-validation was done over a range of bandwidths using only the last 2 steps of Algorithm \ref{Bandwidth_MF_Algo}.
}

\tr {Results for point prediction including bias and mean square error (MSE) over all MB and MF methods are shown in Table \ref{pp_Mudelsee} below.
}

\tr {Results for prediction intervals including CVR, length and standard deviation of
the predicted intervals over all MB and MF methods are shown in Table \ref{bt_Mudelsee} below. The number of bootstrap replications $B$ was set to 1000.}

\tr{
From point-prediction results on this dataset it can be seen that the MF-LLM and LMF-LLM estimators give the best performance.
\tb{The MF-LLM and LMF-LLM estimators also have} the highest coverage probabilities for prediction interval estimation among all estimators that are considered here.
For comparison purposes we have listed the performance of point prediction using the RAMPFIT algorithm outlined in \cite{mudelsee2000ramp} and also used for the speleothem dataset in \cite{fleitmann2003holocene}.} 

\tnr{RAMPFIT introduced by \cite{mudelsee2000ramp} is a popular algorithm used to fit climate data which show transitions such as the speleothem dataset. This algorithm was designed to handle change points in climate time-series and to the best of our knowledge cannot handle arbitrary local stationarity which may be present in data. Hence we chose to use RAMPFIT to compare performance of point prediction versus that obtained using our MB and MF point predictors. \tbl{The MF-LLM-ARMA and LMF-LLM-ARMA  estimators outperform RAMPFIT for point prediction as shown in Table \ref{pp_Mudelsee}. We attribute the superior results of MF-LLM-ARMA and LMF-LLM-ARMA for point prediction and prediction intervals    to the most likely reason that the data is not compatible with the assumption of an additive model.}
RAMPFIT was not originally designed to generate prediction interval estimates hence comparisons of these interval metrics versus those obtained using our MB and MF methods are not provided. The RAMPFIT algorithm is described in Appendix \ref{Appendix_Rampfit}.}

\tr{
For point prediction there is a difference in performance between fitted and predictive residuals which is not the case with the simulation datasets
discussed before. This is} 
\tc{due to} 
\tr{finite sample effects as we use only a small part of the whole speleothem dataset to illustrate the performance
differences between the various estimators. Prediction interval coverage is better using predictive as compared to fitted residuals which is consistent with the results
associated with i.i.d.~regression   \cite{Politis2013}.}

\tbl{As a final point, we consider the practical problem of out-of-sample prediction of the next data point i.e. prediction of  $ Y_{1346}$   using RAMPFIT and our best predictor (MF-LLM-ARMA)  chosen based on in-sample performance. The predicted values using RAMPFIT and MF-LLM-ARMA are nearly the same (which is reassuring), and approximately equal to -0.81.
The $90\%$ prediction interval using MF-LLM is $(-1.165,-0.513 )$; as previously mentioned, RAMPFIT cannot be used to
generate a prediction interval.}

\begin{table}
\centering
\caption{Point Prediction performance for speleothem dataset}
\label{pp_Mudelsee}
\begin{tabular}{|w|w|w|w|}
\hline
Prediction \ Method  & Residual Type & Bias & MSE\\
\hline
{MB-LC} & P &{-5.800e-03} & {4.248e-02}\\
\hline
& F &  {-1.845e-02} & {4.081e-02}\\
\hline
{MB-LL} & P & {1.219e-02}& {4.205e-02}\\
\hline
& F & {1.227e-03} & {3.891e-02}\\
\hline
MF-LC & P  & {-2.755e-02} & {4.006e-02}\\
\hline
& F & {-1.535e-02} & {3.805e-02}\\
\hline
MF-LLH & P & {-2.762e-02} & {3.683e-02}\\
\hline
& F & {-2.141e-02} & {3.925e-02} \\
\hline
MF-LLM & {P} & {-3.776e-03} & {3.513e-02}\\
\hline
& F & {-2.593e-02} & {3.730e-02}\\
\hline
LMF-LC & P &  {-2.602e-02} & {3.959e-02}\\
\hline
& F & {-1.524e-02} & {3.815e-02}\\
\hline
LMF-LLH & P & {-2.672e-02} & {3.682e-02}\\
\hline
& F & {-2.060e-02} & {4.011e-02}\\
\hline
LMF-LLM & P & {5.724e-03} & {3.494e-02}\\
\hline
& F & {-2.702e-02} & {3.643e-02}\\
\hline
MF-LC-ARMA & P & {-2.999e-02} & {4.171e-02} \\
\hline
& F & {-2.058e-02} & {3.874e-02} \\
\hline
MF-LLH-ARMA & P & {-1.8842e-02} & {4.242e-02} \\
\hline
& F & {-1.299e-02} & {3.894e-02} \\
\hline
MF-LLM-ARMA & P & {-3.235e-03} & {3.645e-02} \\
\hline
& F & {-2.077e-02} & {3.427e-02} \\
\hline
LMF-LC-ARMA & P & {-2.718e-02} & {4.143e-02}  \\
\hline
& F & {-2.388e-02}  & {3.953e-02}  \\
\hline
LMF-LLH-ARMA & P & {-1.461e-02} & {4.550e-02}  \\
\hline
& F &  {-1.355e-02} & {4.095e-02}  \\
\hline
LMF-LLM-ARMA & P & {3.538e-03}  & {3.721e-02}  \\
\hline
& F & {-2.174e-02} & {3.550e-02} \\
\hline
\tnb{RAMPFIT} & \tnb{Not Applicable} & \tnb{1.781e-02} & \tnb{3.913e-02}\\
\hline
\end{tabular}
\end{table}

\begin{table}
\centering
\caption{Interval estimation performance using bootstrap for speleothem dataset}
\label{bt_Mudelsee}
\begin{tabular}{| w | w | w | w | w |}
  \hline
  Prediction Method  & Residual Type & CVR & Mean Length & SD Length\\
  \hline
   MB-LC & P & {0.82} & {7.812e-01} &  {2.178e-01} \\
  \hline
  & F &  {0.78} & {5.46e-01} & {1.885e-01} \\
  \hline
 {MB-LL} & P & {0.87} & {8.731e-01} &  {1.970e-01} \\
  \hline
   & F & {0.84} & {7.254e-01} &  {1.689e-01} \\
\hline
  MF-LC & P  & {0.94} & {7.963e-01} & {1.631e-01}\\
  \hline
  & F & {0.84} & {5.076e-01} & {1.525e-01}\\
  \hline
 {MF-LLH} & P & {0.87} & {7.252e-01} &  {1.372e-01}\\
  \hline
  & F & {0.84} & {5.868e-01} &  {1.747e-01} \\
  \hline
   {MF-LLM} & {P} & {0.90} & {7.230e-01} & {1.914e-01} \\
  \hline
  & {F} & {0.89} & {5.788e-01} & {1.774e-01} \\
  \hline
   LMF-LC & P & {0.95} & {7.855e-01} & {1.804e-01} \\
  \hline
  & F & {0.84} & {5.010e-01} & {1.454e-01} \\
  \hline
 LMF-LLH & P & {0.89} & {7.284e-01} & {1.396e-01}\\
  \hline
  & F & {0.81} & {5.568e-01} & {1.613e-01} \\
  \hline
  LMF-LLM & P & {0.90} & {7.397e-01} &  {1.946e-01} \\
  \hline
  & F & {0.89} & {6.145e-01} &  {1.814e-01} \\
  \hline
  MF-LC-ARMA & P & {0.90} & {8.088e-01} & {1.535e-01}\\
  \hline
  & F & {0.86} & {5.754e-01} & {1.665e-01}\\
  \hline
  MF-LLH-ARMA & P & {0.86} & {7.701e-01} & {1.588e-01}\\
  \hline
  & F & {0.80} & {5.759e-01} & {1.911e-01}\\
  \hline
  MF-LLM-ARMA & P & {0.89} & {7.427e-01} & {1.715e-01}\\
  \hline
  & F & {0.86} & {5.819e-01} & {1.973e-01}\\
  \hline
  LMF-LC-ARMA & P & {0.89} & {8.213e-01}  & {1.721e-01}\\
  \hline
  & F & {0.84} & {5.690e-01} & {1.599e-01}\\
  \hline
  LMF-LLH-ARMA & P & {0.87} & {7.783e-01}  & {1.527e-01}\\
  \hline
  & F & {0.78} & {5.772e-01} & {1.916e-01}\\
  \hline
  LMF-LLM-ARMA & P & {0.91} & {7.780e-01}  & {1.818e-01}\\
  \hline
  & F & {0.87} & {6.234e-01} & {2.096e-01}\\
  \hline
  \end{tabular}
\end{table}

%

\graphicspath{{/Users/rumpagiri/Documents/NONPARAMETRIC/model_free/papers/LSTS}}
\DeclareGraphicsExtensions{.png}

{\begin{figure}[!t]
  \centering
  \includegraphics[width=3.5in, height=2.5in]{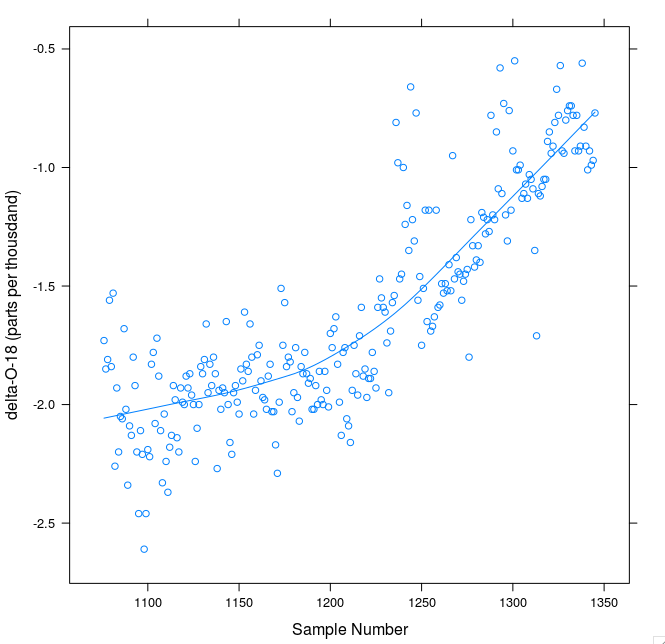}
  \caption{Speleothem data segment used for cross-validation and prediction}
  \label{Mudelsee_short}
\end{figure}
}

\appendix 
\section{Appendix:  Basic Model-free Bootstrap \tr {and Double Bootstrap} Algorithms}
\label{Appendix_Boot}

\tb{This section describes in detail algorithms \ref{MF3short.Algorithm1} and \ref{MF3short.Algorithm2} for the construction of Model-Free and Limit Model-Free algorithms \tor{as described in} \cite{politis2015model}. However note that we also present new algorithms \ref{doublebootmb.Algorithm} and \ref{doublebootmf.Algorithm} to determine bandwidth inside the bootstrap loop for the Model-Based and Model-Free cases.}

Define the 
 {\it  predictive root} to be the error in prediction, i.e., 
\begin{equation}
 Y_{n+1}  - \Pi (   \hat g_{n+1}, \underline{Y}_{n}, \hat F_n) 
\label{MF3short.eq.prederror}
\end{equation}
where $ \Pi(   \hat g_{n+1}, \underline{Y}_{n}, \hat F_n) $
is our chosen point predictor of $Y_{n+1}$, and
$\hat g_{n+1}$ is our estimate of function $  g_{n+1}$ based on the 
data $\underline{Y}_{n}$.

Given bootstrap data $\underline{Y}_{n}^*$ and $Y_{n+1}^*  $, 
  the   bootstrap predictive root  is the error in prediction in the bootstrap world, i.e., 
\begin{equation}
 Y_{n+1}^*  -  \Pi (   \hat g^*_{n+1}, \underline{Y}_{n} , \hat F_n )       
\label{MF3short.eq.bootprederror}
\end{equation}
where
$\hat g^*_{n+1}$ is our estimate of function $  g_{n+1}$ based on the 
bootstrap data $\underline{Y}_{n}^*$.

\begin{Remark} \rm
Note that  eq.~\eqref{MF3short.eq.bootprederror} depends on 
the bootstrap data $\underline{Y}_{n}^*$  only through the estimated 
function $\hat g^*_{n+1}$; both the predictor $\Pi (   \hat g^*_{n+1}, \underline{Y}_{n} , \hat F_n )  $ and the construction of future value $Y_{n+1}^*$
in the sequel are based on the true dataset $\underline{Y}_{n}$ in order
to  give validity to the prediction intervals {\it conditionally} on 
the data $\underline{Y}_{n}$.
\end{Remark} 

\vskip .17in
\begin{Algorithm} {\sc 
Model-free bootstrap for    prediction intervals for
  $ Y_{n+1} $}
\label{MF3short.Algorithm1}
\begin{enumerate}
\item Based on the data $\underline{Y}_n$, estimate the
transformation  $H_n$ and its inverse $H_n^{-1}$
by $\hat H_n$ and  $\hat H_n^{-1}$ respectively.
 In addition, estimate $  g_{n+1}$  by $\hat g_{n+1}$.

\item Use $\hat H_n$ to obtain the transformed
data, i.e.,  $(\varepsilon_1^{(n)}, ..., \varepsilon_n^{(n)})' =\hat H_n (\underline{Y}_n).
$   By construction, the variables 
$\varepsilon_1^{(n)}, ..., \varepsilon_n^{(n)}$
are approximately i.i.d.; let $\hat F_n$ denote their empirical distribution.

\begin{enumerate}

\item Sample randomly (with replacement) the
data $\varepsilon_1^{(n)}, ..., \varepsilon_n^{(n)}$
to create the bootstrap pseudo-data $\varepsilon_1^{\ast }, ..., \varepsilon_n^{\ast}$. 

\item Use the inverse transformation  $\hat H_n^{-1}$ to
create pseudo-data in the $Y$ domain, i.e.,
let $\underline{Y}_n^\ast=(Y_1^\ast ,...,Y_n^\ast )'=
\hat H_n^{-1} (\varepsilon_1^{\ast}, ..., \varepsilon_n^{\ast})$.

\item Calculate a  bootstrap pseudo-response
$Y_{n+1}^*$
as the point
$  \hat g_{n+1} 
( \underline{Y}_n 
,   \varepsilon   )
$ where $\varepsilon $ is drawn randomly from the set
 $(\varepsilon_1^{(n)}, ..., \varepsilon_n^{(n)}) $.

\item Based on the pseudo-data $\underline{Y}_n^\ast$, estimate the
 function  $g_{n+1}$
by    $\hat g_{n+1}^\ast$ respectively.

\item Calculate a   bootstrap root 
replicate  using eq.~(\ref{MF3short.eq.bootprederror}).

\end{enumerate}

\item Steps (a)---(e) in the above should be repeated  a large number
of   times (say $B$ times),
  and the $B$ bootstrap root 
replicates 
should be collected in the form of an empirical distribution whose  
$\alpha$---quantile is denoted by $q_{ }(\alpha)$.
 
\item
A $(1-\alpha)$100\% {\it equal-tailed}
prediction interval 
 for $ Y_{n+1} $ is given by
\begin{equation}
[\Pi +q_{ }(\alpha/2), \ 
\Pi + q_{ }(1-\alpha/2)] 
\label{MF3short.eq.predint2.4root}
\end{equation}
where $\Pi $ is short-hand for $
\Pi (   \hat g_{n+1}, \underline{Y}_{n},  \hat F_n)$. 
 
\end{enumerate}
\end{Algorithm}
\vskip .173in
\noindent
   Sometimes, the empirical distribution 
$\hat F_n$ converges to a limit distribution $F$
that is of known form (perhaps after   
estimating a finite-dimensional parameter). Using it instead of the empirical
$\hat F_n$ results into the 
Limit Model-Free (LMF) resampling algorithm that is given below.
Note that now the point predictor $\Pi$ is no more  a function  of 
$\hat F_n$ but of $F$.
Hence, the LMF  predictive  root is denoted by 
\begin{equation}
   Y_{n+1}  - \Pi (   \hat g_{n+1}, \underline{Y}_{n},  
F)  
\label{MF3short.eq.prederrorH}
\end{equation}
  whose distribution
  can be approximated by that of the LMF bootstrap  predictive root 
\begin{equation}
   Y_{n+1}^*  -  \Pi (  \hat g^*_{n+1}, \underline{Y}_{n} 
,  F)  . 
\label{MF3short.eq.bootprederrorH}
\end{equation}

\vskip .20in
\begin{Algorithm} 
\label{MF3short.Algorithm2}
{\sc Limit Model-free (LMF) bootstrap for    prediction intervals for 
$ Y_{n+1} $}  

\begin{enumerate}
\item Based on the data $\underline{Y}_n$, estimate the
transformation  $H_n$ and its inverse $H_n^{-1}$
by $\hat H_n$ and  $\hat H_n^{-1}$ respectively.
 In addition, estimate $  g_{n+1}$  by $\hat g_{n+1}$.

\item 

\begin{enumerate}

\item 
Generate bootstrap pseudo-data $\varepsilon_1^{\ast }, ..., \varepsilon_n^{\ast}$
in an i.i.d.~manner from $F$.  

\item Use the inverse transformation  $\hat H_n^{-1}$ to
create pseudo-data in the $Y$ domain, i.e.,
let $\underline{Y}_n^\ast=(Y_1^\ast ,...,Y_n^\ast )'=
\hat H_n^{-1} (\varepsilon_1^{\ast}, ..., \varepsilon_n^{\ast})$.

\item Calculate a  bootstrap pseudo-response
$Y_{n+1}^*$
as the point
$  \hat g_{n+1} 
( \underline{Y}_n 
 ,   \varepsilon   )
$ where $\varepsilon $ is a random draw from distribution $F$.

\item Based on the pseudo-data $\underline{Y}_n^\ast$, estimate the
 function  $g_{n+1}$
by    $\hat g_{n+1}^\ast$ respectively.

\item Calculate a   bootstrap root 
replicate  using eq.~(\ref{MF3short.eq.bootprederrorH}).

\end{enumerate}

\item Steps (a)---(e) in the above should be repeated  a large number
of   times (say $B$ times),
  and the $B$ bootstrap root 
replicates 
should be collected in the form of an empirical distribution whose  
$\alpha$---quantile is denoted by $q_{ }(\alpha)$.
 
\item
A $(1-\alpha)$100\% {\it equal-tailed}
prediction interval 
 for $ Y_{n+1} $ is given by
\begin{equation}
[\Pi +q_{ }(\alpha/2)   ,  \ 
\Pi + q_{ }(1-\alpha/2)    ] 
\label{MF3short.eq.predint2.4root}
\end{equation}
where $\Pi $ is short-hand for $
\Pi (   \hat g_{n+1}, \underline{Y}_{n} ,F)$.

\end{enumerate}
\end{Algorithm}


\noindent
\tr
{
Both Model-Based and Model-Free bootstrap algorithms enable the construction of prediction intervals for a pre-determined nominal coverage level.
Point-prediction can use the bandwidth $b$ determined by the respective cross-validation procedures outlined for the MB and MF cases
in Sections \ref{NSTS.sec.trend} and \ref{NSTS.sec.MF.cv} respectively. However to prevent under or overcoverage \tb{with respect to} the nominal level during
calculation of prediction intervals we recommend a double bootstrap procedure to accurately set the bandwidth $b'$ inside the bootstrap loop
which uses the resampled residuals from point prediction in both the MB and MF cases. The algorithms 
\ref{doublebootmb.Algorithm} and \ref {doublebootmf.Algorithm} below enable the determination of this adjusted bandwidth $b'$.
}

\newpage

\begin{Algorithm}
\label{doublebootmb.Algorithm}

\tr {{\sc MB double bootstrap for  bandwidth in bootstrap loop}  }

\begin{enumerate}
\item 
\tr {
Based on the data $Y_1,\ldots, Y_n$ and the bandwidth $b$ based on model-based cross-validation, calculate the estimators $\check \mu (\cdot) $ and $\check \sigma (\cdot) $,
and the `residuals'  $\check W_1, \ldots, \check W_n$ using
model   \eqref{NSTS.eq.model hetero}.
}

\item
\tr {
Fit the AR($p$) model~\eqref{NSTS.eq.AR} to the series $\check W_1, \ldots, \check W_n$ (with $p$ selected by AIC minimization), and obtain the  Yule-Walker estimators $\hat \phi_1 , \ldots , \hat \phi_p$, and the error proxies 
$$\check V_t=\check W_{t} -
  \hat \phi_1 \check W_{t-1} - \cdots -  \hat \phi_p \check W_{t-p  }
\ \ \mbox{for} \ \  
\ \  t=p+b+1,\ldots, n.$$}

\item  
\tr {
Let $\check V_{t}^*$ for $t=  1,\ldots, n,n+1$ be drawn randomly with
replacement from the set  $\{\check {\check {V_t}}$ for 
$  t=p+b+1,\ldots, n\} $
where  
$\check {\check {V_t}} =\check V_{t}  - (n-p-b)^{-1}\sum_{i=p+b+1}^n  \check V_i$.
Let $I$ be a random variable drawn from a discrete uniform distribution
on the values  
$\  p+b,p+b+1,\ldots, n\ $,
and define the bootstrap initial conditions 
$ \check W_t^* =\check W_{t+I}$ for $t= -p+1,\ldots,0$.
Then, create the bootstrap data $\check W_1^*,\ldots,\check W_n^* $ via the  AR recursion 
 $$\check W_t^*= \hat \phi_1 \check W_{t-1}^* + \cdots +  \hat \phi_p \check W_{t-p  }^* + \check V_{t}^*\ \ \mbox{for} \ \  t= 1,\ldots, (n+1). $$ 
 This is the first bootstrap loop.}
 
 \item
 \tr {Create the bootstrap pseudo-series
$Y_1^*,\ldots, Y_{n+1}^*$ by the formula
$$ Y_t^*= \check \mu (t) +\check \sigma (t) \check W_t^*
\ \ \mbox{for} \ \  t= 1,\ldots, (n+1). $$}

\item 
\tr {
Based on the data $Y_1^*,\ldots, Y_n^*$ (first $n$ values only) and the bandwidth $b$ based on model-based cross-validation,
calculate the estimators $\check \mu (\cdot)^* $ and $\check \sigma (\cdot)^* $,
and the `residuals'  $ W_1^*, \ldots,  W_n^*$ using
model   \eqref{NSTS.eq.model hetero}.
}

\item
\tr {
Fit the AR($p$) model~\eqref{NSTS.eq.AR} to the series $ W_1^*, \ldots, W_n^*$ (with $p$ selected by AIC minimization), and obtain the  Yule-Walker estimators $\hat \phi_1^* , \ldots , \hat \phi_p^*$, and the error proxies 
$$\check V_t^*= W_{t}^* -
  \hat \phi_1^* W_{t-1}^* - \cdots -  \hat \phi_p^* W_{t-p}^*
\ \ \mbox{for} \ \  
\ \  t=p+b+1,\ldots, n.$$ 
}

\item  
\begin{enumerate}

\item
\tr {
Let $\check V_{t}^{**}$ for $t=  1,\ldots, n,n+1$ be drawn randomly with
replacement from the set  $\{\check {\check {V_t}}^*$ for 
$  t=p+b+1,\ldots, n\} $
where  
$\check {\check {V_t}}^* =\check V_{t}^*  - (n-p-b)^{-1}\sum_{i=p+b+1}^n  \check V_i^*$.
Let $I$ be a random variable drawn from a discrete uniform distribution
on the values  
$\  p+b,p+b+1,\ldots, n\ $,
and define the bootstrap initial conditions 
$ \check W_t^{**} = W_{t+I}^{*}$ for $t= -p+1,\ldots,0$.
Then, create the bootstrap data $\check W_1^{**},\ldots,\check W_n^{**} $ via the  AR recursion 
 $$\check W_t^{**}= \hat \phi_1 W_{t-1}^{**} + \cdots +  \hat \phi_p W_{t-p  }^{**} + \check V_{t}^{**}\ \ \mbox{for} \ \  t= 1,\ldots, (n+1). $$ 
 This is the second bootstrap loop.}
 
 \item 
 \tr {Create the bootstrap pseudo-series
$Y_1^{**},\ldots, Y_n^{**}$ by the formula
$$ Y_t^{**}= \check \mu (t)^* +\check \sigma (t)^* \check W_t^{**}
\ \ \mbox{for} \ \  t= 1,\ldots, n. $$} 

\item 
\tr {
Re-calculate the estimators $\check \mu^{**} (\cdot) $ and $\check \sigma^{**} (\cdot) $  from the bootstrap data
$Y_1^*,\ldots, Y_n^*$. 
The bootstrap estimators $\check \mu^{**} (\cdot) $ and $\check \sigma^{**} (\cdot) $
are based on a bandwidth value $b'$ which is different from the bandwidth $b$ obtained by model-based cross-validation.
This gives rises to new
bootstrap residuals $\check W_1^{**}, \ldots, \check W_n^{**}$ on which an AR($p$) model is again fitted yielding the
bootstrap  Yule-Walker estimators $\hat \phi_1^{**} , \ldots , \hat \phi_p^{**}$.
}

\item 
\tr {
Calculate the bootstrap predictor
 $$\Pi^{**}= \check \mu ^{**} (n+1) +\check \sigma^{**} (n+1)
\left[\hat \phi_1^{**}  W_{ n}^* + \ldots + \hat \phi_p^{**} W_{n-p+1 }^*\right].$$}

\item 
\tr {
Calculate a  bootstrap future value 
$$Y_{n+1}^{**} = \check \mu^*  (n+1) +\check \sigma^* (n+1) W_{n+1 }^{**}$$
where again 
$ W_{n+1 }^{**}=\hat \phi_1^* W_{n}^*  + \cdots +  \hat \phi_p^* W_{n-p+1}^* + \check V_{n+1}^{**}$ uses the original values
$( W_{ n}^* , \ldots ,W_{n-p+1 }^*)$; recall that $\check V_{n+1}^{**}$ has  already been generated in step (a) above.
}
 
\item 
\tr {
Calculate the bootstrap root replicate  $Y_{n+1}^{**}-\Pi^{**}$.
}

\end{enumerate}

\item 
\tr {Steps (a)---(f) in the above are repeated  a large number
of   times (say $C$ times),
  and the $C$ bootstrap root replicates 
are collected in the form of an empirical distribution whose  
$\alpha$--quantile is denoted by $q_{ }(\alpha)$.
}

\item
\tr {
Finally, a $(1-\alpha)$100\% {\it equal-tailed}
prediction interval 
 for $ Y_{n+1}^* $ ($n$th value of $\underline{Y}_{n+1}^\ast$) is given by
\begin{equation}
[\Pi^* +q_{ }(\alpha/2), \ 
\Pi^* + q_{ }(1-\alpha/2)] .
\label{NSTS.eq.predint2.4rootdouble}
\end{equation} 
}

\tr {
Here $\Pi^*$ is given by:
}
\tr {
\begin{equation}
\Pi^*= \check \mu^* (n+1) +\check \sigma^* (n+1)
\left[ \hat \phi_1^* W_{n}^* + \cdots + \hat \phi_p^* W_{n-p+1}^* \right] 
\label{NSTS.eq.predictorPistar}
\end{equation}
where     
$\hat \phi_1^* , \ldots , \hat \phi_p^*$ are the Yule-Walker estimators of
$  \phi_1 , \ldots ,   \phi_p$ appearing in eq.~\eqref{NSTS.eq.AR}.
}

\item
\tr {
Steps (3)--(9) in the above should be repeated a large number of times (say $B$ times)
to obtain $B$ values of $Y_{n+1}^*$ and their corresponding $(1-\alpha)$100\% {\it equal-tailed}
prediction intervals as outlined by Step (9) above. This can then be used to calculate
a coverage probability (CVR) for various values of the second bootstrap loop ($C$ iterations) bandwidth
$b'$ while keeping the bandwidth $b$ of the outer bootstrap loop ($B$ iterations) fixed to what was
obtained from cross-validation. The value of $b'$ that gives the target CVR can be used as
the bandwidth for the bootstrap loop in Algorithm \ref{NSTS.AlgorithmMB}.
}

\end{enumerate}

\end{Algorithm}

\begin{Algorithm}
\label{doublebootmf.Algorithm}
\tr {{\sc MF double bootstrap for  bandwidth in bootstrap loop}  }

\begin{enumerate}

\item
\tr{ 
Based on the data $\underline{Y}_n$ and the bandwidth $b$ obtained from
model-free cross-validation, estimate the
transformation  $H_n$ and its inverse $H_n^{-1}$
by $\hat H_n$ and  $\hat H_n^{-1}$ respectively.
In addition, estimate $  g_{n+1}$  by $\hat g_{n+1}$.
}

\item 
\tr {
Use $\hat H_n$ to obtain the transformed
data, i.e.,  $(\varepsilon_1^{(n)}, ..., \varepsilon_n^{(n)})' =\hat H_n (\underline{Y}_n).
$   By construction, the variables 
$\varepsilon_1^{(n)}, ..., \varepsilon_n^{(n)}$
are approximately i.i.d.
}

\item
\tr {
Sample randomly (with replacement) the
data $\varepsilon_1^{(n)}, ..., \varepsilon_n^{(n)}$
to create the bootstrap pseudo-data $\varepsilon_1^{\ast }, ..., \varepsilon_{n+1}^{\ast}$.
This is the first bootstrap loop.
}

\item
\tr{
Use the inverse transformation  $\hat H_n^{-1}$ and the bandwidth $b$ from model-free cross-validation
to create pseudo-data in the $Y$ domain, i.e., let
$\underline{Y}_{n+1}^\ast=(Y_1^\ast ,...,Y_{n+1}^\ast )'=\hat H_n^{-1} (\varepsilon_1^{\ast}, ..., \varepsilon_{n+1}^{\ast})$.}

\item
\tr {
Based on the data $\underline{Y}_n^\ast$ (first $n$ values only) and the bandwidth $b$ obtained from
model-free cross-validation, estimate the
transformation  $H_n^*$ and its inverse $H_n^{*-1}$
by $\hat H_n^*$ and  $\hat H_n^{*-1}$ respectively.
In addition, estimate $  g_{n+1}$  by $\hat g_{n+1}^{*}$.
}

\item
\tr{
Use $\hat H_n^*$ to obtain the transformed
data, i.e.,  $(\varepsilon_1^{*(n)}, ..., \varepsilon_n^{*(n)})' =\hat H_n^{*} (\underline{Y}^*_n).
$   By construction, the variables 
$\varepsilon_1^{*(n)}, ..., \varepsilon_n^{*(n)}$
are approximately i.i.d; let $\hat F_n^*$ denote their empirical distribution.
}
\begin{enumerate}
\item
\tr{
Sample randomly (with replacement) the
data $\varepsilon_1^{*(n)}, ..., \varepsilon_n^{*(n)}$
to create the bootstrap pseudo-data $\varepsilon_1^{**(n)}, ..., \varepsilon_n^{**(n)}.$
This is the second bootstrap loop.
}
\item
\tr{
Use the inverse transformation  $\hat H_n^{*-1}$ and a bandwidth $b^{'}$ (different from $b$ found from model-free cross-validation)
to create pseudo-data in the $Y$ domain, i.e., let
$\underline{Y}_n^{**}=(Y_1^{**} ,...,Y_n^{**} )'=\hat H_n^{*-1} (\varepsilon_1^{**}, ..., \varepsilon_n^{**})$.
}

\item 
\tr{
Calculate a  bootstrap pseudo-response
$Y_{n+1}^{**}$
as the point
$  \hat g_{n+1}^* 
( \underline{Y}_n^* 
 ,   \varepsilon^{*}   )
 $ where $\varepsilon^* $ is drawn randomly from the set
 $(\varepsilon_1^{*(n)}, ..., \varepsilon_n^{*(n)}) $.
}

\item 
\tr{
Based on the pseudo-data $\underline{Y}_n^{**}$ and bandwidth $b'$, estimate the
 function  $g_{n+1}$ by    $\hat g_{n+1}^{**}$ respectively.
 }
 
 \item 
 \tr {
 Calculate a   bootstrap root  replicate  using
 \begin{equation}
   Y_{n+1}^{**}  -  \Pi (  \hat g^{**}_{n+1}, \underline{Y}_{n}^* 
,  \hat F_n^*)  . 
\label{MF3short.eq.bootprederrorHpseudo}
\end{equation}}
\end{enumerate}

\item 
\tr{
Steps (a)---(e) in the above should be repeated  a large number
of   times (say $C$ times),
  and the $C$ bootstrap root 
replicates 
should be collected in the form of an empirical distribution whose  
$\alpha$---quantile is denoted by $q_{ }(\alpha)$.
}

\item
\tr {
A $(1-\alpha)$100\% {\it equal-tailed}
prediction interval 
 for $ Y_{n+1}^* $ ($n$th value of $\underline{Y}_{n+1}^\ast$) is given by
\begin{equation}
[\Pi^* +q_{ }(\alpha/2), \ 
\Pi^* + q_{ }(1-\alpha/2)] 
\label{MF3short.eq.predint2.4rootpseudo}
\end{equation}
where $\Pi^* $ is short-hand for $
\Pi (   \hat g_{n+1}^*, \underline{Y}_{n}^*,  \hat F_n^*)$. 
}

\item
\tr {
Steps (3)--(8) in the above should be repeated a large number of times (say $B$ times)
to obtain $B$ values of $Y_{n+1}^*$ and their corresponding $(1-\alpha)$100\% {\it equal-tailed}
prediction intervals as outlined by Step (8) above. This can then be used to calculate
a coverage probability (CVR) for various values of the second bootstrap loop ($C$ iterations) bandwidth
$b'$ while keeping the bandwidth $b$ of the outer bootstrap loop ($B$ iterations) fixed to what was
obtained from cross-validation. The value of $b'$ that gives the target CVR can be used as
the bandwidth for the bootstrap loop in Algorithms \ref{MF3short.Algorithm1} and
\ref{MF3short.Algorithm2}.
}

\end{enumerate}
\end{Algorithm}

\section{\tc{Appendix: RAMPFIT algorithm for analyzing climate data with transitions}  }
\label{Appendix_Rampfit}

\tc{
The RAMPFIT algorithm
} 
\tnr{which can handle uneven time-spacing in observations}
\tc{
was proposed by \cite{mudelsee2000ramp} for performing regression on climate data which shows transitions such as the speleothem dataset considered in this paper.} 
\tnr{However RAMPFIT was not originally designed to handle arbitrary local stationarity which may be present in data.} 
\tc
{Here we briefly outline the steps in RAMPFIT used to obtain point prediction estimates which are used for comparison with their Model-Based and Model-Free counterparts.}
\smallskip

\noindent
\tnr{Define $x(i) = X(t(i))$ where ($X_t, t \in {\bf R}$) is an underlying continuous-time stochastic process.}
\tc{
For a time series $x(i)$ measured at times $t(i), i = 1,\ldots,n,$ the model under consideration is \cite{mudelsee2000ramp}:}

\begin{equation}
x(i) = x_{fit}(i) + \epsilon(i)
\label{RAMPFIT}
\end{equation}

\bigskip
\noindent
\tc{
It is assumed that the errors $\epsilon(i)$ are heteroskedastic and are distributed as $N(0, \sigma(i)^2)$}.

\noindent
\bigskip
\tc{
The fitted model is a ramp function as defined below:}

\begin{align}
x_{fit}(t) = \begin{cases}
                         \  x1, \ \hspace{49 mm} for \  t \leq t1,\\
                         \  x1 + (t-t1)(x2-x1)/(t2-t1), \ for \ t1 \leq t \leq t2,\\
                         \  x2, \ \hspace{49 mm} for \ t \geq t2
                \end{cases} \label{ramp_equation}
\end{align}

\tc{
Here $t1$ and $t2$ denote the start and end of the ramp and $x1$, $x2$ denote the corresponding values at those points. The regression model is fitted to data \{$t(i)$, $x(i)$\}$_{i=1}^{n}$  by minimizing the weighted sum of squares as given below:}
\begin{equation}
SSQW(t1,x1,t2,x2) = \sum_{i=1}^{n} \frac {[x(i) - x_{fit}(i)]^2} {\sigma(i)^2}
\label{SSQW_eqn}
\end{equation}

\tc{
Owing to the non-differentiabilities at $t1$ and $t2$, RAMPFIT does a search over a range of values supplied for these 2 values and chooses the values ($\hat t1, \hat x1, \hat t2, \hat x2$) for which the $SSQW$ is minimum. In addition since $\sigma(i)$ is not known an initial guess of this is supplied to the algorithm following which the $\sigma(i)$ values are recalculated from the obtained residuals. The estimates ($\hat t1, \hat x1, \hat t2, \hat x2$) are then regenerated. These steps are repeated till MSE values of point prediction converge.}

\bigskip
\noindent
\tc{The full algorithm is described below:}

\bigskip
\begin{Algorithm} {
RAMPFIT REGRESSION
\label{rampfit_regression}
}
\label{rampfit_iteration}
\begin{enumerate} 
\item \tnb{Set initial estimate of $\sigma(i) = i$ with $i =1,\ldots,n$}
\item Set search ranges [$t1_{min}$, $t1_{max}$] and [$t2_{min}$,$t2_{max}$] for values of $t1$ and $t2$
\item Calculate SSQW using (\ref{ramp_equation}) and (\ref{SSQW_eqn}) over this grid of $t1$ and $t2$ values; denote a typical point in this grid as ($\bar t1$, $\bar t2$)
\item \tnr{Determine $(\hat t1, \hat x1, \hat t2, \hat x2)$ = argmin $[SSQW(\bar t1, \hat x1, \bar t2, \hat x2)]$ and obtain $x_{fit}$}
\item Calculate residuals $e(i) = x(t(i)) - x_{fit}(t(i))$
\item \tnr{Re-estimate the variance $\sigma(i)$ from $e(i)$ using k-nearest-neighbour smoothing}
\item Repeat steps (2) to (6) above till MSE values converge.
\end{enumerate}
\end{Algorithm}


 
\vskip .1in
\clearpage
\noindent 
{\bf Acknowledgements} \\
This research was partially supported by NSF grants DMS 12-23137 and DMS 16-13026. The authors would like to acknowledge the Pacific Research Platform, NSF Project ACI-1541349 and Larry Smarr (PI, Calit2 at UCSD) for providing the computing infrastructure used in this project. Many thanks are also due to Richard Davis and Stathis Paparoditis for their helpful comments. The authors would also like to thank Dr. Manfred Mudelsee for kindly sharing the FORTRAN code for the RAMPFIT algorithm used for analyzing the speleothem dataset used in this paper.

\bibliography{srinjoy_stats}

\begin{thebibliography}{}

\bibitem [\protect \citeauthoryear {%
Brockwell%
\ \BBA {} Davis%
}{%
Brockwell%
\ \BBA {} Davis%
}{%
{\protect \APACyear {1991}}%
}]{%
brockwell2013time}
\APACinsertmetastar {%
brockwell2013time}%
\begin{APACrefauthors}%
Brockwell, P\BPBI J.%
\BCBT {}\ \BBA {} Davis, R\BPBI A.%
\end{APACrefauthors}%
\unskip\
\newblock
\APACrefYear{1991}.
\newblock
\APACrefbtitle {Time series: theory and methods} {Time series: theory and
  methods}\ (\PrintOrdinal{Second}\ \BEd).
\newblock
\APACaddressPublisher{}{Springer, New York}.
\PrintBackRefs{\CurrentBib}

\bibitem [\protect \citeauthoryear {%
Dahlhaus%
}{%
Dahlhaus%
}{%
{\protect \APACyear {2012}}%
}]{%
dahlhaus2012locally}
\APACinsertmetastar {%
dahlhaus2012locally}%
\begin{APACrefauthors}%
Dahlhaus, R.%
\end{APACrefauthors}%
\unskip\
\newblock
\APACrefYearMonthDay{2012}{}{}.
\newblock
{\BBOQ}\APACrefatitle {Locally stationary processes} {Locally stationary
  processes}.{\BBCQ}
\newblock
\BIn{} T\BPBI S.~Rao\ \BOthers {.}\ (\BEDS), \APACrefbtitle {Handbook of
  statistics} {Handbook of statistics}\ (\BVOL~30, \BPG~351-412).
\newblock
\APACaddressPublisher{}{Elsevier}.
\PrintBackRefs{\CurrentBib}

\bibitem [\protect \citeauthoryear {%
Dahlhaus%
\ \protect \BOthers {.}}{%
Dahlhaus%
\ \protect \BOthers {.}}{%
{\protect \APACyear {1997}}%
}]{%
dahlhaus1997fitting}
\APACinsertmetastar {%
dahlhaus1997fitting}%
\begin{APACrefauthors}%
Dahlhaus, R.%
\BCBT {}\ \BOthersPeriod {.}
\end{APACrefauthors}%
\unskip\
\newblock
\APACrefYearMonthDay{1997}{}{}.
\newblock
{\BBOQ}\APACrefatitle {Fitting time series models to nonstationary processes}
  {Fitting time series models to nonstationary processes}.{\BBCQ}
\newblock
\APACjournalVolNumPages{The Annals of Statistics}{25}{1}{1--37}.
\PrintBackRefs{\CurrentBib}

\bibitem [\protect \citeauthoryear {%
Das%
\ \BBA {} Politis%
}{%
Das%
\ \BBA {} Politis%
}{%
{\protect \APACyear {2017}}%
}]{%
das2017nonparametric}
\APACinsertmetastar {%
das2017nonparametric}%
\begin{APACrefauthors}%
Das, S.%
\BCBT {}\ \BBA {} Politis, D\BPBI N.%
\end{APACrefauthors}%
\unskip\
\newblock
\APACrefYearMonthDay{2017}{}{}.
\newblock
{\BBOQ}\APACrefatitle {Nonparametric estimation of the conditional distribution
  at regression boundary points} {Nonparametric estimation of the conditional
  distribution at regression boundary points}.{\BBCQ}
\newblock
\APACjournalVolNumPages{arXiv preprint arXiv:1704.00674}{}{}{}.
\PrintBackRefs{\CurrentBib}

\bibitem [\protect \citeauthoryear {%
Dowla%
, Paparoditis%
\BCBL {}\ \BBA {} Politis%
}{%
Dowla%
\ \protect \BOthers {.}}{%
{\protect \APACyear {2013}}%
}]{%
dowla2013local}
\APACinsertmetastar {%
dowla2013local}%
\begin{APACrefauthors}%
Dowla, A.%
, Paparoditis, E.%
\BCBL {}\ \BBA {} Politis, D\BPBI N.%
\end{APACrefauthors}%
\unskip\
\newblock
\APACrefYearMonthDay{2013}{}{}.
\newblock
{\BBOQ}\APACrefatitle {Local block bootstrap inference for trending time
  series} {Local block bootstrap inference for trending time series}.{\BBCQ}
\newblock
\APACjournalVolNumPages{Metrika}{76}{6}{733--764}.
\PrintBackRefs{\CurrentBib}

\bibitem [\protect \citeauthoryear {%
Dowla~A.%
\ \BBA {} \mbox{Politis D.N}%
}{%
Dowla~A.%
\ \BBA {} \mbox{Politis D.N}%
}{%
{\protect \APACyear {2003}}%
}]{%
dowla2003locally}
\APACinsertmetastar {%
dowla2003locally}%
\begin{APACrefauthors}%
Dowla~A., {\relax Paparoditis E}.%
\BCBT {}\ \BBA {} \mbox{Politis D.N}.%
\end{APACrefauthors}%
\unskip\
\newblock
\APACrefYearMonthDay{2003}{}{}.
\newblock
{\BBOQ}\APACrefatitle {Locally stationary processes and the local block
  bootstrap} {Locally stationary processes and the local block
  bootstrap}.{\BBCQ}
\newblock
\BIn{} M\BPBI G.~Akritas\ \BBA {} D\BPBI N.~Politis\ (\BEDS), \APACrefbtitle
  {Recent advances and trends in nonparametric statistics} {Recent advances and
  trends in nonparametric statistics}\ (\BPG~437-444).
\newblock
\APACaddressPublisher{}{Elsevier}.
\PrintBackRefs{\CurrentBib}

\bibitem [\protect \citeauthoryear {%
Fan%
\ \BBA {} Gijbels%
}{%
Fan%
\ \BBA {} Gijbels%
}{%
{\protect \APACyear {1996}}%
}]{%
fan1996local}
\APACinsertmetastar {%
fan1996local}%
\begin{APACrefauthors}%
Fan, J.%
\BCBT {}\ \BBA {} Gijbels, I.%
\end{APACrefauthors}%
\unskip\
\newblock
\APACrefYear{1996}.
\newblock
\APACrefbtitle {Local polynomial modelling and its applications: monographs on
  statistics and applied probability} {Local polynomial modelling and its
  applications: monographs on statistics and applied probability}\ (\BVOL~66).
\newblock
\APACaddressPublisher{}{CRC Press, Boca Raton}.
\PrintBackRefs{\CurrentBib}

\bibitem [\protect \citeauthoryear {%
Fan%
\ \BBA {} Yao%
}{%
Fan%
\ \BBA {} Yao%
}{%
{\protect \APACyear {2007}}%
}]{%
fan2007nonlinear}
\APACinsertmetastar {%
fan2007nonlinear}%
\begin{APACrefauthors}%
Fan, J.%
\BCBT {}\ \BBA {} Yao, Q.%
\end{APACrefauthors}%
\unskip\
\newblock
\APACrefYear{2007}.
\newblock
\APACrefbtitle {Nonlinear time series: nonparametric and parametric methods}
  {Nonlinear time series: nonparametric and parametric methods}.
\newblock
\APACaddressPublisher{}{Springer, New York}.
\PrintBackRefs{\CurrentBib}

\bibitem [\protect \citeauthoryear {%
Fleitmann%
\ \protect \BOthers {.}}{%
Fleitmann%
\ \protect \BOthers {.}}{%
{\protect \APACyear {2003}}%
}]{%
fleitmann2003holocene}
\APACinsertmetastar {%
fleitmann2003holocene}%
\begin{APACrefauthors}%
Fleitmann, D.%
, Burns, S\BPBI J.%
, Mudelsee, M.%
, Neff, U.%
, Kramers, J.%
, Mangini, A.%
\BCBL {}\ \BBA {} Matter, A.%
\end{APACrefauthors}%
\unskip\
\newblock
\APACrefYearMonthDay{2003}{}{}.
\newblock
{\BBOQ}\APACrefatitle {Holocene forcing of the Indian monsoon recorded in a
  stalagmite from southern Oman} {Holocene forcing of the indian monsoon
  recorded in a stalagmite from southern oman}.{\BBCQ}
\newblock
\APACjournalVolNumPages{Science}{300}{5626}{1737--1739}.
\PrintBackRefs{\CurrentBib}

\bibitem [\protect \citeauthoryear {%
Hall%
, Wolff%
\BCBL {}\ \BBA {} Yao%
}{%
Hall%
\ \protect \BOthers {.}}{%
{\protect \APACyear {1999}}%
}]{%
Hall1999}
\APACinsertmetastar {%
Hall1999}%
\begin{APACrefauthors}%
Hall, P.%
, Wolff, R\BPBI C.%
\BCBL {}\ \BBA {} Yao, Q.%
\end{APACrefauthors}%
\unskip\
\newblock
\APACrefYearMonthDay{1999}{}{}.
\newblock
{\BBOQ}\APACrefatitle {Methods for estimating a conditional distribution
  function} {Methods for estimating a conditional distribution
  function}.{\BBCQ}
\newblock
\APACjournalVolNumPages{Journal of the American Statistical
  Association}{94}{445}{154--163}.
\PrintBackRefs{\CurrentBib}

\bibitem [\protect \citeauthoryear {%
Hansen%
}{%
Hansen%
}{%
{\protect \APACyear {2004}}%
}]{%
hansen2004nonparametric}
\APACinsertmetastar {%
hansen2004nonparametric}%
\begin{APACrefauthors}%
Hansen, B\BPBI E.%
\end{APACrefauthors}%
\unskip\
\newblock
\APACrefYearMonthDay{2004}{}{}.
\newblock
{\BBOQ}\APACrefatitle {Nonparametric estimation of smooth conditional
  distributions} {Nonparametric estimation of smooth conditional
  distributions}.{\BBCQ}
\newblock
\APACjournalVolNumPages{Unpublished paper: Department of Economics, University
  of Wisconsin}{}{}{}.
\PrintBackRefs{\CurrentBib}

\bibitem [\protect \citeauthoryear {%
H{\"a}rdle%
\ \BBA {} Vieu%
}{%
H{\"a}rdle%
\ \BBA {} Vieu%
}{%
{\protect \APACyear {1992}}%
}]{%
hardle1992kernel}
\APACinsertmetastar {%
hardle1992kernel}%
\begin{APACrefauthors}%
H{\"a}rdle, W.%
\BCBT {}\ \BBA {} Vieu, P.%
\end{APACrefauthors}%
\unskip\
\newblock
\APACrefYearMonthDay{1992}{}{}.
\newblock
{\BBOQ}\APACrefatitle {Kernel regression smoothing of time series} {Kernel
  regression smoothing of time series}.{\BBCQ}
\newblock
\APACjournalVolNumPages{Journal of Time Series Analysis}{13}{3}{209--232}.
\PrintBackRefs{\CurrentBib}

\bibitem [\protect \citeauthoryear {%
Jentsch%
\ \BBA {} Politis%
}{%
Jentsch%
\ \BBA {} Politis%
}{%
{\protect \APACyear {2015}}%
}]{%
jentsch2015covariance}
\APACinsertmetastar {%
jentsch2015covariance}%
\begin{APACrefauthors}%
Jentsch, C.%
\BCBT {}\ \BBA {} Politis, D\BPBI N.%
\end{APACrefauthors}%
\unskip\
\newblock
\APACrefYearMonthDay{2015}{}{}.
\newblock
{\BBOQ}\APACrefatitle {Covariance matrix estimation and linear process
  bootstrap for multivariate time series of possibly increasing dimension}
  {Covariance matrix estimation and linear process bootstrap for multivariate
  time series of possibly increasing dimension}.{\BBCQ}
\newblock
\APACjournalVolNumPages{The Annals of Statistics}{43}{3}{1117--1140}.
\PrintBackRefs{\CurrentBib}

\bibitem [\protect \citeauthoryear {%
Kim%
\ \BBA {} Cox%
}{%
Kim%
\ \BBA {} Cox%
}{%
{\protect \APACyear {1996}}%
}]{%
kim1996bandwidth}
\APACinsertmetastar {%
kim1996bandwidth}%
\begin{APACrefauthors}%
Kim, T\BPBI Y.%
\BCBT {}\ \BBA {} Cox, D\BPBI D.%
\end{APACrefauthors}%
\unskip\
\newblock
\APACrefYearMonthDay{1996}{}{}.
\newblock
{\BBOQ}\APACrefatitle {Bandwidth selection in kernel smoothing of time series}
  {Bandwidth selection in kernel smoothing of time series}.{\BBCQ}
\newblock
\APACjournalVolNumPages{Journal of Time Series Analysis}{17}{1}{49--63}.
\PrintBackRefs{\CurrentBib}

\bibitem [\protect \citeauthoryear {%
Kreiss%
, Paparoditis%
\BCBL {}\ \BBA {} Politis%
}{%
Kreiss%
\ \protect \BOthers {.}}{%
{\protect \APACyear {2011}}%
}]{%
kreiss2011range}
\APACinsertmetastar {%
kreiss2011range}%
\begin{APACrefauthors}%
Kreiss, J\BHBI P.%
, Paparoditis, E.%
\BCBL {}\ \BBA {} Politis, D\BPBI N.%
\end{APACrefauthors}%
\unskip\
\newblock
\APACrefYearMonthDay{2011}{}{}.
\newblock
{\BBOQ}\APACrefatitle {On the range of validity of the autoregressive sieve
  bootstrap} {On the range of validity of the autoregressive sieve
  bootstrap}.{\BBCQ}
\newblock
\APACjournalVolNumPages{The Annals of Statistics}{39}{4}{2103--2130}.
\PrintBackRefs{\CurrentBib}

\bibitem [\protect \citeauthoryear {%
Li%
\ \BBA {} Racine%
}{%
Li%
\ \BBA {} Racine%
}{%
{\protect \APACyear {2007}}%
}]{%
li2007nonparametric}
\APACinsertmetastar {%
li2007nonparametric}%
\begin{APACrefauthors}%
Li, Q.%
\BCBT {}\ \BBA {} Racine, J\BPBI S.%
\end{APACrefauthors}%
\unskip\
\newblock
\APACrefYear{2007}.
\newblock
\APACrefbtitle {Nonparametric econometrics: theory and practice} {Nonparametric
  econometrics: theory and practice}.
\newblock
\APACaddressPublisher{}{Princeton University Press, Princeton}.
\PrintBackRefs{\CurrentBib}

\bibitem [\protect \citeauthoryear {%
Masry%
\ \BBA {} Tj{\o}stheim%
}{%
Masry%
\ \BBA {} Tj{\o}stheim%
}{%
{\protect \APACyear {1995}}%
}]{%
masry1995nonparametric}
\APACinsertmetastar {%
masry1995nonparametric}%
\begin{APACrefauthors}%
Masry, E.%
\BCBT {}\ \BBA {} Tj{\o}stheim, D.%
\end{APACrefauthors}%
\unskip\
\newblock
\APACrefYearMonthDay{1995}{}{}.
\newblock
{\BBOQ}\APACrefatitle {Nonparametric Estimation and Identification of Nonlinear
  ARCH Time Series Strong Convergence and Asymptotic Normality: Strong
  Convergence and Asymptotic Normality} {Nonparametric estimation and
  identification of nonlinear arch time series strong convergence and
  asymptotic normality: Strong convergence and asymptotic normality}.{\BBCQ}
\newblock
\APACjournalVolNumPages{Econometric theory}{11}{2}{258--289}.
\PrintBackRefs{\CurrentBib}

\bibitem [\protect \citeauthoryear {%
McMurry%
\ \BBA {} Politis%
}{%
McMurry%
\ \BBA {} Politis%
}{%
{\protect \APACyear {2010}}%
}]{%
mcmurry2010banded}
\APACinsertmetastar {%
mcmurry2010banded}%
\begin{APACrefauthors}%
McMurry, T\BPBI L.%
\BCBT {}\ \BBA {} Politis, D\BPBI N.%
\end{APACrefauthors}%
\unskip\
\newblock
\APACrefYearMonthDay{2010}{}{}.
\newblock
{\BBOQ}\APACrefatitle {Banded and tapered estimates for autocovariance matrices
  and the linear process bootstrap} {Banded and tapered estimates for
  autocovariance matrices and the linear process bootstrap}.{\BBCQ}
\newblock
\APACjournalVolNumPages{Journal of Time Series Analysis}{31}{6}{471--482}.
\PrintBackRefs{\CurrentBib}

\bibitem [\protect \citeauthoryear {%
McMurry%
\ \BBA {} Politis%
}{%
McMurry%
\ \BBA {} Politis%
}{%
{\protect \APACyear {2015}}%
}]{%
mcmurry2015high}
\APACinsertmetastar {%
mcmurry2015high}%
\begin{APACrefauthors}%
McMurry, T\BPBI L.%
\BCBT {}\ \BBA {} Politis, D\BPBI N.%
\end{APACrefauthors}%
\unskip\
\newblock
\APACrefYearMonthDay{2015}{}{}.
\newblock
{\BBOQ}\APACrefatitle {High-dimensional autocovariance matrices and optimal
  linear prediction} {High-dimensional autocovariance matrices and optimal
  linear prediction}.{\BBCQ}
\newblock
\APACjournalVolNumPages{Electronic Journal of Statistics}{9}{1}{753--788}.
\PrintBackRefs{\CurrentBib}

\bibitem [\protect \citeauthoryear {%
Mudelsee%
}{%
Mudelsee%
}{%
{\protect \APACyear {2000}}%
}]{%
mudelsee2000ramp}
\APACinsertmetastar {%
mudelsee2000ramp}%
\begin{APACrefauthors}%
Mudelsee, M.%
\end{APACrefauthors}%
\unskip\
\newblock
\APACrefYearMonthDay{2000}{}{}.
\newblock
{\BBOQ}\APACrefatitle {Ramp function regression: a tool for quantifying climate
  transitions} {Ramp function regression: a tool for quantifying climate
  transitions}.{\BBCQ}
\newblock
\APACjournalVolNumPages{Computers \& Geosciences}{26}{3}{293--307}.
\PrintBackRefs{\CurrentBib}

\bibitem [\protect \citeauthoryear {%
Mudelsee%
}{%
Mudelsee%
}{%
{\protect \APACyear {2014}}%
}]{%
mudelsee2014climate}
\APACinsertmetastar {%
mudelsee2014climate}%
\begin{APACrefauthors}%
Mudelsee, M.%
\end{APACrefauthors}%
\unskip\
\newblock
\APACrefYear{2014}.
\newblock
\APACrefbtitle {Climate time series analysis: classical statistical and
  bootstrap methods} {Climate time series analysis: classical statistical and
  bootstrap methods}.
\newblock
\APACaddressPublisher{}{Springer Science and Business Media}.
\PrintBackRefs{\CurrentBib}

\bibitem [\protect \citeauthoryear {%
Pan%
\ \BBA {} Politis%
}{%
Pan%
\ \BBA {} Politis%
}{%
{\protect \APACyear {2016}}%
}]{%
pan2016bootstrap}
\APACinsertmetastar {%
pan2016bootstrap}%
\begin{APACrefauthors}%
Pan, L.%
\BCBT {}\ \BBA {} Politis, D\BPBI N.%
\end{APACrefauthors}%
\unskip\
\newblock
\APACrefYearMonthDay{2016}{}{}.
\newblock
{\BBOQ}\APACrefatitle {Bootstrap prediction intervals for Markov processes}
  {Bootstrap prediction intervals for markov processes}.{\BBCQ}
\newblock
\APACjournalVolNumPages{Computational Statistics \& Data
  Analysis}{100}{}{467--494}.
\PrintBackRefs{\CurrentBib}

\bibitem [\protect \citeauthoryear {%
Paparoditis%
\ \BBA {} Politis%
}{%
Paparoditis%
\ \BBA {} Politis%
}{%
{\protect \APACyear {2002}}%
}]{%
paparoditis2002local}
\APACinsertmetastar {%
paparoditis2002local}%
\begin{APACrefauthors}%
Paparoditis, E.%
\BCBT {}\ \BBA {} Politis, D\BPBI N.%
\end{APACrefauthors}%
\unskip\
\newblock
\APACrefYearMonthDay{2002}{}{}.
\newblock
{\BBOQ}\APACrefatitle {Local block bootstrap} {Local block bootstrap}.{\BBCQ}
\newblock
\APACjournalVolNumPages{Comptes Rendus Mathematiques}{335}{11}{959--962}.
\PrintBackRefs{\CurrentBib}

\bibitem [\protect \citeauthoryear {%
Politis%
}{%
Politis%
}{%
{\protect \APACyear {2001}}%
}]{%
politis2001nonparametric}
\APACinsertmetastar {%
politis2001nonparametric}%
\begin{APACrefauthors}%
Politis, D\BPBI N.%
\end{APACrefauthors}%
\unskip\
\newblock
\APACrefYearMonthDay{2001}{}{}.
\newblock
{\BBOQ}\APACrefatitle {On nonparametric function estimation with infinite-order
  flat-top kernels} {On nonparametric function estimation with infinite-order
  flat-top kernels}.{\BBCQ}
\newblock
\BIn{} {\relax Ch. A. Charalambides}\ \BOthers {.}\ (\BEDS), \APACrefbtitle
  {Probability and Statistical Models with applications} {Probability and
  statistical models with applications}\ (\BPG~469-483).
\newblock
\APACaddressPublisher{}{Chapman and Hall/CRC: Boca Raton}.
\PrintBackRefs{\CurrentBib}

\bibitem [\protect \citeauthoryear {%
Politis%
}{%
Politis%
}{%
{\protect \APACyear {2013}}%
}]{%
Politis2013}
\APACinsertmetastar {%
Politis2013}%
\begin{APACrefauthors}%
Politis, D\BPBI N.%
\end{APACrefauthors}%
\unskip\
\newblock
\APACrefYearMonthDay{2013}{}{}.
\newblock
{\BBOQ}\APACrefatitle {Model-free model-fitting and predictive distributions}
  {Model-free model-fitting and predictive distributions}.{\BBCQ}
\newblock
\APACjournalVolNumPages{Test}{22}{2}{183--221}.
\PrintBackRefs{\CurrentBib}

\bibitem [\protect \citeauthoryear {%
Politis%
}{%
Politis%
}{%
{\protect \APACyear {2015}}%
}]{%
politis2015model}
\APACinsertmetastar {%
politis2015model}%
\begin{APACrefauthors}%
Politis, D\BPBI N.%
\end{APACrefauthors}%
\unskip\
\newblock
\APACrefYear{2015}.
\newblock
\APACrefbtitle {Model-Free Prediction and Regression} {Model-free prediction
  and regression}.
\newblock
\APACaddressPublisher{}{Springer, New York}.
\PrintBackRefs{\CurrentBib}

\bibitem [\protect \citeauthoryear {%
Politis%
\ \BBA {} Romano%
}{%
Politis%
\ \BBA {} Romano%
}{%
{\protect \APACyear {1994}}%
}]{%
politis1994stationary}
\APACinsertmetastar {%
politis1994stationary}%
\begin{APACrefauthors}%
Politis, D\BPBI N.%
\BCBT {}\ \BBA {} Romano, J\BPBI P.%
\end{APACrefauthors}%
\unskip\
\newblock
\APACrefYearMonthDay{1994}{}{}.
\newblock
{\BBOQ}\APACrefatitle {The stationary bootstrap} {The stationary
  bootstrap}.{\BBCQ}
\newblock
\APACjournalVolNumPages{Journal of the American Statistical
  association}{89}{428}{1303--1313}.
\PrintBackRefs{\CurrentBib}

\bibitem [\protect \citeauthoryear {%
Priestley%
}{%
Priestley%
}{%
{\protect \APACyear {1965}}%
}]{%
priestley1965evolutionary}
\APACinsertmetastar {%
priestley1965evolutionary}%
\begin{APACrefauthors}%
Priestley, M\BPBI B.%
\end{APACrefauthors}%
\unskip\
\newblock
\APACrefYearMonthDay{1965}{}{}.
\newblock
{\BBOQ}\APACrefatitle {Evolutionary spectra and non-stationary processes}
  {Evolutionary spectra and non-stationary processes}.{\BBCQ}
\newblock
\APACjournalVolNumPages{Journal of the Royal Statistical Society. Series B
  (Methodological)}{}{}{204--237}.
\PrintBackRefs{\CurrentBib}

\bibitem [\protect \citeauthoryear {%
Priestley%
}{%
Priestley%
}{%
{\protect \APACyear {1988}}%
}]{%
priestley1988non}
\APACinsertmetastar {%
priestley1988non}%
\begin{APACrefauthors}%
Priestley, M\BPBI B.%
\end{APACrefauthors}%
\unskip\
\newblock
\APACrefYear{1988}.
\newblock
\APACrefbtitle {Non-linear and non-stationary time series analysis} {Non-linear
  and non-stationary time series analysis}.
\newblock
\APACaddressPublisher{}{Academic Press, London}.
\PrintBackRefs{\CurrentBib}

\bibitem [\protect \citeauthoryear {%
Rosenblatt%
}{%
Rosenblatt%
}{%
{\protect \APACyear {1952}}%
}]{%
rosenblatt1952remarks}
\APACinsertmetastar {%
rosenblatt1952remarks}%
\begin{APACrefauthors}%
Rosenblatt, M.%
\end{APACrefauthors}%
\unskip\
\newblock
\APACrefYearMonthDay{1952}{}{}.
\newblock
{\BBOQ}\APACrefatitle {Remarks on a multivariate transformation} {Remarks on a
  multivariate transformation}.{\BBCQ}
\newblock
\APACjournalVolNumPages{The Annals of Mathematical
  Statistics}{23}{3}{470--472}.
\PrintBackRefs{\CurrentBib}

\bibitem [\protect \citeauthoryear {%
Samorodnitsky%
\ \BBA {} Taqqu%
}{%
Samorodnitsky%
\ \BBA {} Taqqu%
}{%
{\protect \APACyear {1994}}%
}]{%
samorodnitsky1994stable}
\APACinsertmetastar {%
samorodnitsky1994stable}%
\begin{APACrefauthors}%
Samorodnitsky, G.%
\BCBT {}\ \BBA {} Taqqu, M\BPBI S.%
\end{APACrefauthors}%
\unskip\
\newblock
\APACrefYearMonthDay{1994}{}{}.
\newblock
\APACrefbtitle {Stable Non-Gaussian Random Processes: Stochastic Models with
  Infinite Variance (Stochastic Modeling Series).} {Stable non-gaussian random
  processes: Stochastic models with infinite variance (stochastic modeling
  series).}
\newblock
\APACaddressPublisher{}{Chapman and Hall/CRC Press}.
\PrintBackRefs{\CurrentBib}

\bibitem [\protect \citeauthoryear {%
Tong%
}{%
Tong%
}{%
{\protect \APACyear {2011}}%
}]{%
tong2011threshold}
\APACinsertmetastar {%
tong2011threshold}%
\begin{APACrefauthors}%
Tong, H.%
\end{APACrefauthors}%
\unskip\
\newblock
\APACrefYearMonthDay{2011}{}{}.
\newblock
{\BBOQ}\APACrefatitle {Threshold models in time series analysis—---30 years
  on} {Threshold models in time series analysis—---30 years on}.{\BBCQ}
\newblock
\APACjournalVolNumPages{Statistics and its Interface}{4}{2}{107--118}.
\PrintBackRefs{\CurrentBib}

\bibitem [\protect \citeauthoryear {%
Zhou%
\ \BBA {} Wu%
}{%
Zhou%
\ \BBA {} Wu%
}{%
{\protect \APACyear {2009}}%
}]{%
zhou2009local}
\APACinsertmetastar {%
zhou2009local}%
\begin{APACrefauthors}%
Zhou, Z.%
\BCBT {}\ \BBA {} Wu, W\BPBI B.%
\end{APACrefauthors}%
\unskip\
\newblock
\APACrefYearMonthDay{2009}{}{}.
\newblock
{\BBOQ}\APACrefatitle {Local linear quantile estimation for nonstationary time
  series} {Local linear quantile estimation for nonstationary time
  series}.{\BBCQ}
\newblock
\APACjournalVolNumPages{The Annals of Statistics}{37}{5B}{2696--2729}.
\PrintBackRefs{\CurrentBib}

\bibitem [\protect \citeauthoryear {%
Zhou%
\ \BBA {} Wu%
}{%
Zhou%
\ \BBA {} Wu%
}{%
{\protect \APACyear {2010}}%
}]{%
zhou2010simultaneous}
\APACinsertmetastar {%
zhou2010simultaneous}%
\begin{APACrefauthors}%
Zhou, Z.%
\BCBT {}\ \BBA {} Wu, W\BPBI B.%
\end{APACrefauthors}%
\unskip\
\newblock
\APACrefYearMonthDay{2010}{}{}.
\newblock
{\BBOQ}\APACrefatitle {Simultaneous inference of linear models with time
  varying coefficients} {Simultaneous inference of linear models with time
  varying coefficients}.{\BBCQ}
\newblock
\APACjournalVolNumPages{Journal of the Royal Statistical Society: Series B
  (Statistical Methodology)}{72}{4}{513--531}.
\PrintBackRefs{\CurrentBib}

\end{thebibliography}
\bibliographystyle{apacite}
 
 \end{document}